\documentclass[aps,prd,amsmath,amsfonts,a4paper,reprint,preprintnumbers,twocolumn,nofootinbib,superscriptaddress]{revtex4-1}

\usepackage{hyperref}
\usepackage{natbib}
\usepackage{amsmath}
\usepackage{amsfonts}
\usepackage{amssymb}
\usepackage{color}
\usepackage{bm}

\usepackage{graphicx}

\usepackage{feynmp}
\usepackage[labelfont={bf,scriptsize},textfont={scriptsize},justification=RaggedRight,format=hang]{caption,subfig}

\usepackage[T1]{fontenc}
\usepackage[charter]{mathdesign}
\RequirePackage{ifpdf}

\ifpdf
\else 
\fi

\newcommand{\vect}[1]{\mathbf{{#1}}}
\newcommand{\e}[1]{\mathrm{e}^{{#1}}}

\newcommand{\Mp}{M_{\mathrm{P}}}

\newcommand{\etal}{et~al.}

\newcommand{\fNL}{f_{\mathrm{NL}}}
\newcommand{\tauNL}{\tau_{\mathrm{NL}}}
\newcommand{\gNL}{g_{\mathrm{NL}}}

\newcommand{\smoothscale}{L}
\newcommand{\coscale}{\mu}
\newcommand{\varN}{N'}
\newcommand{\adlimit}{\mathrm{AL}}

\newcommand{\Sinit}{\mathcal{S}}
\newcommand{\Ainit}{\mathcal{A}}

\newcommand{\refractiveindex}{\nu}

\newcommand{\Hessian}{\mathcal{H}}
\newcommand{\kindex}[1]{{#1}'}
\newcommand{\isoindex}[1]{\mathbf{#1}}
\newcommand{\isosigma}{\sigma^{\text{iso}}}

\DeclareMathOperator{\pathorder}{\mathcal{P}}
\DeclareMathOperator{\isopartial}{\mathcal{D}}
\DeclareMathOperator{\adpartial}{\mathcal{D}_{\sigma}}

\renewcommand{\geq}{\geqslant}
\DeclareMathOperator{\Or}{O}

\DeclareMathOperator{\tr}{tr}

\renewcommand{\d}{\mathrm{d}}

\newcommand{\para}[1]{\par\vspace{2mm}\noindent\textbf{{#1}}.---}

\newcommand{\subpara}[1]{\par\vspace{1mm}\noindent\emph{{#1}}.---}

\begin{document}

	\title{Inflationary perturbation theory is
	geometrical optics in phase space}

	\author{David Seery}
	\email{D.Seery@sussex.ac.uk}
	\affiliation{Astronomy Centre, University of Sussex, Brighton BN1 9QH,
	United Kingdom}

	\author{David J. Mulryne}
	\email{D.Mulryne@qmul.ac.uk}
	\affiliation{Astronomy Unit, School of Mathematical Sciences,
	Queen Mary, University of London,
	Mile End Road, London E1 4NS, United Kingdom}

	\author{Jonathan Frazer}
	\email{J.Frazer@sussex.ac.uk}
	\affiliation{Astronomy Centre, University of Sussex, Brighton BN1 9QH,
	United Kingdom}
	
	\author{Raquel H. Ribeiro}
	\email{R.Ribeiro@damtp.cam.ac.uk}
	\affiliation{Department of Applied Mathematics and Theoretical Physics,
	Centre for Mathematical Sciences, University of
	Cambridge, Wilberforce Road,
	Cambridge CB3 0WA, United Kingdom}
	
	\begin{abstract}
		A pressing problem in comparing inflationary models with
		observation is the accurate calculation of correlation functions.
		One approach is to evolve them
		using ordinary differential
		equations (``transport equations''),
		analogous to the Schwinger--Dyson hierarchy of
		\emph{in--out} quantum field theory.
		We extend this approach to the complete set of momentum space
		correlation functions.
		A formal solution can be obtained using
		raytracing
		techniques adapted from
		geometrical optics.
		We reformulate inflationary
		perturbation theory in this language,
		and show that raytracing reproduces
		the familiar ``$\delta N$'' Taylor expansion.
		Our method produces ordinary differential equations which
		allow the Taylor coefficients to be computed efficiently.
		We use raytracing methods to
		express the gauge transformation between field fluctuations
		and the curvature perturbation, $\zeta$, in geometrical terms.
		Using these results we
		give
		a compact expression for the nonlinear gauge-transform
		part of $\fNL$ in terms of
		the principal curvatures of uniform energy-density hypersurfaces
		in field space.
	\end{abstract}
	
	\maketitle
	
	\begin{fmffile}{optics}

	\section{Introduction}
	\label{sec:introduction}
	
	Our current theories of the early universe are stochastic.
	They do not predict
	a definite \emph{state} today:
	rather, their predictions are statistical.
	To compare these predictions with observation it must usually
	be supposed that we are in some sense typical.
	This brings two challenges.
	First, what is typical under some circumstances may be atypical
	under others. Therefore we must be precise about the type of observer
	of which we are a typical representative.
	This leads to the ``measure problem,''
	about which we have nothing new to say.
	In this paper we are concerned with the second challenge:
	after fixing a class of observers,
	to estimate the observables typically measured by
	its members.

	Inflation is the most common early-universe
	paradigm for which we would like to compute observables.
	In this context
	we usually take ourselves to be ordinary observers
	of the fluctuations produced on approach to a
	fixed vacuum.
	The challenge is to calculate the typical stochastic properties
	of these fluctuations.
		
	The most important fluctuation generated by inflation
	is
	the primordial density perturbation, $\zeta$.
	Correlations in the
	temperature and polarization
	anisotropies
	of the microwave background
	are inherited
	from $\zeta$
	and provide a clean probe of its statistical character.
	Therefore,
	both present-day constraints~\cite{Komatsu:2010fb}
	and
	the imminent arrival of high quality
	microwave-background data~\cite{:2006uk,*Ade:2011ah}
	make
	accurate estimates of its statistical properties
	a pressing issue.
	Meanwhile, large surveys of the cosmological density field
	will provide information about its properties
	on complementary, smaller scales \cite{Desjacques:2010jw}.
	To compare this abundance of data to models we require an
	efficient tool with which to
	estimate the $n$-point functions
	$\langle \zeta^n \rangle$.
	
	Taking $\zeta$ to be synthesized from the fluctuation of
	one or more light scalar fields during an inflationary era,
	several computational schemes exist which enable
	the $n$-point functions to be studied.
	Many of these schemes employ some variant of the
	\emph{separate universe picture}
	\cite{Starobinsky:1986fxa,*Lyth:1984gv,*Sasaki:1995aw,
	*Salopek:1990jq,*Sasaki:1998ug,*Wands:2000dp}.
	Taking $H$ to be the Hubble parameter,
	this asserts that---when smoothed on some physical scale,
	$\smoothscale$,
	much larger than the horizon scale, so that
	$\smoothscale/H^{-1} \gg 1$---%
	the average evolution of each $\smoothscale$-sized patch can be computed
	using the \emph{background} equations of motion
	and initial conditions taken from smoothed quantities local to the
	patch.
	Working from a Taylor expansion in the initial conditions
	for each
	patch, Lyth \& Rodr\'{\i}guez
	showed how this assumption could be turned into a practical
	algorithm for calculating correlation functions \cite{Lyth:2005fi}.
	This ``$\delta N$ method'' has become the most popular way to
	explore the predictions of specific models, both
	analytically and numerically,
	and has developed a large literature of its own.
	The principal difficulty arises when calculating the coefficients
	of the Taylor expansion,
	sometimes called the ``$\delta N$ coefficients.''
	We shall discuss this difficulty in~\S\ref{sec:flow-deltaN}.
	
	Alternative approaches exist.
	Rigopoulos, Shellard \& van Tent
	\cite{Rigopoulos:2004gr,*Rigopoulos:2005xx}
	evolved each correlation
	function using a Langevin equation.
	Yokoyama, Suyama \&~Tanaka~%
	\cite{Yokoyama:2007uu,*Yokoyama:2007dw,*Yokoyama:2008by}
	decomposed each $\delta N$ coefficient into components which
	could be computed using ordinary differential equations.
	Later, a systematic method to obtain `transport' equations for
	the entire hierarchy of
	correlation functions
	(rather than simply the $\delta N$ coefficients) was
	introduced~\cite{Mulryne:2009kh,*Mulryne:2010rp}.
	A more longstanding approach
	uses the methods of traditional cosmological perturbation
	theory (``CPT'') to produce ``transfer matrices''
	\cite{Amendola:2001ni}.
	This has recently been revived
	by a number of
	authors~\cite{GrootNibbelink:2001qt,Lalak:2007vi,Peterson:2010np,
	*Peterson:2010mv,*Peterson:2011yt,Achucarro:2010da,Avgoustidis:2011em}.
	Numerical approaches have been employed
	by Lehners \& Renaux-Petel \cite{Lehners:2009ja},
	Ringeval \cite{Ringeval:2007am,*Martin:2006rs}, and
	Huston \&~Malik \cite{Huston:2009ac,*Huston:2011vt}.
	
	The relationship of
	these different methods to each other
	has not always been clear.
	Nor is it always obvious how to relate the approximations
	employed by each technique.
	In this paper we study the connections between many of these
	approaches
	using the formalism
	of Elliston~{\etal}~\cite{Elliston:2011dr}.
	This is a statistical interpretation of the separate universe picture.
	In what follows we briefly summarize the construction.
	(See also Ref.~\cite{Mulryne:2009kh,*Mulryne:2010rp}.)
	
	\para{The separate universe approximation as
	statistical mechanics}%
	Fix a large spacetime box of comoving side $\coscale$
	containing the region of interest. The scale $\coscale$ 
	should be much larger than the
	separate universe scale,
	requiring $\coscale \gg \smoothscale$,
	but not superexponentially larger
	\cite{Boubekeur:2005fj,*Lyth:2006gd,*Seery:2010kh}.
	After smoothing on the scale
	$\smoothscale$, the fields within the large box
	pick out an ensemble or cloud of
	$N \sim (\coscale/\smoothscale)^3$
	points in the classical phase space.
	The condition that $\coscale/\smoothscale$
	is not \emph{superexponentially} large means that
	the typical diameter of the cloud will be
	roughly of
	order the quantum scatter $\langle \delta \phi^2 \rangle^{1/2}
	\sim H$.
	Because $N$ is still large, $N \gg 1$,
	it is convenient to
	describe the ensemble by
	an occupation probability $\rho$ on phase space.%
		\footnote{To be certain that we are
		estimating only the observables measured by a typical
		observer living within a single terminal vacuum,
		we should demand that $\rho$ has support only
		on points whose orbits eventually converge
		in some neighbourhood of that vacuum.
		This requires that all horizon volumes reheat almost surely
		in the same minimum.
		If some horizon volumes reheat in different minima then
		the resulting correlation functions are not measurable
		by a local observer who sees only a single vacuum.}
	The correlation functions of $\zeta$ on the scale $\smoothscale$
	are then determined by the classical statistical mechanics
	of this ensemble,
	which is encoded in the Boltzmann equation.
	
	In familiar applications of statistical mechanics, the evolution
	of the ensemble may be complicated. Small-scale interactions
	scatter members of the cloud between orbits on phase space,
	represented by the collisional term in the Boltzmann equation.
	However, the separate universe assumption
	requires causality to suppress those interactions which would
	be required for scattering between orbits.
	Therefore the evolution is trivial.
	Each point in phase space is assigned an occupation
	probability by the initial conditions,
	which is conserved along its orbit.
	All that is required is a mapping of initial conditions
	to the final state,
	which is obtained by carrying the initial conditions
	along the phase space flow generated by the underlying
	theory.
	It follows that the Boltzmann equation
	can be integrated using the method of characteristics.

\begin{figure*}
		\begin{fmfgraph*}(250,90)
 			\fmfcmd{%
 				style_def double_arrow expr p =
 				cdraw p;
 				cfill (harrow (reverse p, 1));
 				cfill (harrow (p, 1));
 				enddef;}
 			\fmfpen{thin}
			\fmfipair{a,b,c}				
			\fmfipair{e,f,g}				
			\fmfipair{starta,startb,startc}	
			\fmfipair{enda,endb,endc}		
			\fmfipair{mida,midc}			
			\fmfipair{jstart,jend}			
			\fmfiequ{a}{(0,0.3h)}
			\fmfiequ{b}{(0.15w,0.1h)}
			\fmfiequ{c}{(0.2w,0)}
			\fmfi{plain,foreground=blue}{a .. b .. c}
			\fmfiequ{e}{(0.7w,h)}
			\fmfiequ{f}{(0.8w,0.85h)}
			\fmfiequ{g}{(w,0.7h)}
			\fmfi{plain,foreground=blue}{e .. f .. g}
			\fmfiequ{starta}{point 0.25*length(a .. b .. c) of (a .. b .. c)}
			\fmfiequ{startb}{point 0.4*length(a .. b .. c) of (a .. b .. c)}
			\fmfiequ{startc}{point 0.6*length(a .. b .. c) of (a .. b .. c)}
			\fmfiequ{enda}{point 0.1*length(e .. f .. g) of (e .. f .. g)}
			\fmfiequ{endb}{point 0.55*length(e .. f .. g) of (e .. f .. g)}
			\fmfiequ{endc}{point 0.9*length(e .. f .. g) of (e .. f .. g)}
			\fmfiequ{mida}{starta shifted (0.12w,0.1h)}
			\fmfiequ{midc}{startc shifted (0.1w,0.12h)}
			\fmfi{fermion}{starta .. mida .. enda}
			\fmfi{fermion,foreground=red,label={characteristic curves},label.side=right,label.dist=0.1w}{startb .. endb}
			\fmfi{fermion}{startc .. midc .. endc}
			\fmfiequ{jstart}{enda shifted (0.01w,0.05h)}
			\fmfiequ{jend}{endb shifted (0.01w,0.05h)}
			\fmfi{double_arrow,label={Jacobi field},label.side=left}{jstart .. jend}
			\fmfiv{label={final surface},label.angle=180}{e}
			\fmfiv{label={initial surface}}{a}
			\fmfiv{label={conserved density $\rho$},decoration.shape=circle,decoration.size=0.05h,decoration.filled=full}{startc}
			\fmfiv{label={conserved density $\rho$},label.angle=45,decoration.shape=circle,decoration.size=0.05h,decoration.filled=full}{endc}
		\end{fmfgraph*}
		\caption{Jacobi fields. Characteristic curves are labelled by
		arrows, and the red characteristic is the fiducial
		curve.
		A conserved probability density
		is dragged along the flow.
		At any point, the Jacobi fields span the space infinitesimal
		displacements to neighbouring characteristics.\label{fig:jacobi-fields}}
	\end{figure*}
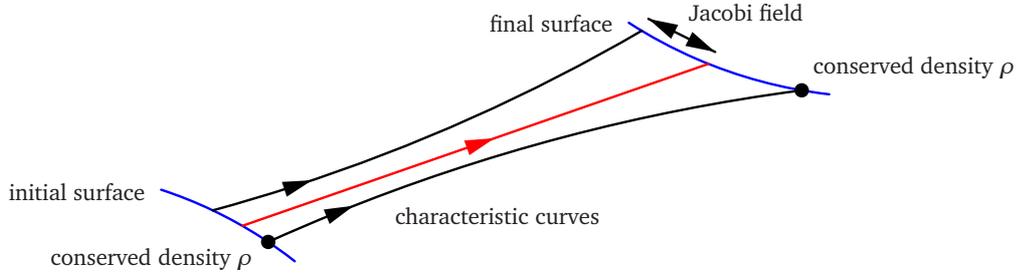

	A similar conclusion applies to the correlation functions of interest,
	$\langle \zeta^n \rangle$.
	These probe information
	about the distribution function over the cloud,
	giving a weighted average over many characteristics.
	Alternatively, if the cloud has only a small phase-space diameter,
	we can exchange information about the entire set
	of characteristics
	for the details of a \emph{single} fiducial characteristic
	and a description of how nearby characteristics separate from it.
	In differential geometry this description is provided by
	the apparatus of Jacobi fields;
	see Fig.~\ref{fig:jacobi-fields}.
	We shall see that
	the differing
	implementations of the separate universe approximation
	can be understood
	as alternative methods to compute these Jacobi fields.
	
	In applications we are frequently interested in correlation
	functions associated with mixed scales, rather than a single
	scale $L$.
	To do so we construct multiple ensembles associated with different
	smoothing scales.
	The separate universe approximation couples the evolution of
	all these
	ensembles in a specific way, which we
	describe in \S\ref{sec:transport}.
	
	\para{Outline}%
	In this paper we
	develop and refine the
	statistical-mechanical
	interpretation of the separate universe picture
	summarized above.
	Because the final distribution of occupation probabilities
	is an \emph{image}
	generated by dragging along the phase space flow,
	it can be calculated
	in precisely the same way that
	geometrical optics enables us to calculate the image generated
	by a source of light rays.
	In \S\ref{sec:geo-optics} we show that, at least within the
	slow-roll approximation,
	this parallel is exact; the scalar field equation
	can be interpreted as the eikonal equation for a light ray
	in a medium with varying refractive index---%
	or equivalently as Huygens' equation for a wavefront.
	
	In \S\S\ref{sec:jacobi}--\ref{sec:adiabatic}
	we introduce the idea of Jacobi fields
	and explore their connection with the ``adiabatic limit,''
	in which all isocurvature modes decay and
	the curvature perturbation becomes conserved.
	Such limits are important because an inflationary model
	is predictive \emph{on its own} only if the flow enters such a region
	\cite{Polarski:1994rz,GarciaBellido:1995qq,*Wands:1996kb,Langlois:1999dw,
	Elliston:2011dr}.
	Jacobi fields
	are familiar from the description of congruences of light
	rays in general relativity \cite{Hawking:1973uf,deFelice:1990hu}.
	In this case, as shown in Fig.~\ref{fig:jacobi-fields},
	they describe fluctuations between the
	$\smoothscale$-sized patches
	which make up the ensemble.
	Their evolution enables the adiabatic and isocurvature
	modes to be tracked.
	In particular, decay of isocurvature modes means decay of
	the corresponding Jacobi fields,
	which occurs when the bundle of trajectories undergoes
	focusing.
	
	In \S\ref{sec:transport} we use these ideas to develop
	evolution (``transport'') equations for each correlation function,
	and in \S\ref{sec:evolution} we show that the Jacobi fields
	can be used to formally integrate the system of transport
	equations.
	This gives a practical method to identify regions where the
	flow becomes adiabatic.
	The analysis can begin from either the separate universe
	principle or traditional cosmological perturbation theory.
	As a by-product, our formal solution demonstrates that the
	transport equations are equivalent
	to the Taylor expansion algorithm introduced by
	Lyth \& Rodr\'{\i}guez.

	In \S\ref{sec:flow-deltaN}
	we use this solution to derive
	a closed set of differential equations
	for the Taylor coefficients,
	and
	in \S\ref{sec:shape}
	we explain how the
	transport equations can be manipulated
	to obtain evolution equations for the
	coefficients of each momentum ``shape''.
	These shapes will be an important diagnostic
	tool when comparing inflationary models to observation
	\cite{Ribeiro:2011ax, *Battefeld:2011ut}.
	Together with the transport hierarchy of \S\ref{sec:transport},
	the equations of
	\S\S\ref{sec:flow-deltaN}--\ref{sec:shape}
	represent the principal results of this paper.
	Either set can be used to obtain the correlation functions of
	a given theory, and we discuss their comparative
	advantages.

	In \S\ref{sec:connections}
	we give more
	a more general discussion
	of the relationship between the transport equations and
	other formulations of perturbation theory.
	
	In \S\ref{sec:gauge} we
	specialize to the slow-roll approximation and
	use ray-tracing methods to derive the
	gauge transformation between field fluctuations and the curvature
	perturbation, $\zeta$.
	As a result, we obtain
	the gauge transformation in terms of geometrical
	quantities---in particular, the extrinsic
	curvature of constant density hypersurfaces.
	We separate the
	gauge contribution to $\fNL$
	into a number of effects,
	corresponding to these geometrical quantities.
	For some models, we
	show that the largest of these
	can be attributed to
	a strong \emph{relative} enhancement 
	of the power in isocurvature fluctuations.
	We briefly
	discuss what conclusions can be drawn regarding
	the asymptotic magnitude of $|\fNL|$.
	
	Finally, we provide a brief summary of our results in
	\S\ref{sec:summary}.
	
	\para{Notation and conventions}%
	We use units in which $c = \hbar = 1$, and work in terms of the
	reduced Planck mass, $\Mp^{-2} = 8 \pi G$.
	We use a number of index conventions
	which are introduced
	in the text.
	See especially the paragraph \emph{Index convention}
	on p.~\pageref{para:index-convention},
	and the discussion of primed indices below
	Eq.~\eqref{eq:k-deviation} on p.~\pageref{primed-index-convention}.
	
	\section{Geometrical optics in phase space}
	\label{sec:geo-optics}

	Throughout this paper,
	our discussion will
	apply to an inflationary phase which
	can
	be described by a collection of
	canonical
	scalar fields $\phi_\alpha$ coupled to Einstein gravity.
	We initially use
	Greek labels $\alpha$, $\beta$, {\ldots}, to label
	the different species of fields.
	The action for this system is
	\begin{equation}
		S = \frac{1}{2} \int \d^4 x \; \sqrt{-g} \left(
			\Mp^2 R - \partial_a \phi_\alpha \partial^a \phi_\alpha
			- 2 V
		\right) ,
		\label{eq:einstein-hilbert}
	\end{equation}
	where $V = V(\phi^\alpha)$ is an interaction potential depending only
	on the scalar fields,
	and indices $a$, $b$, {\ldots}, run over space time dimensions.
	We take the background geometry to be flat
	Friedmann--Robertson--Walker
	with scale factor $a(t)$.
	
	\subsection{Slow-roll approximation: rays on field space}
	\label{sec:slow-roll}
	
	In this subsection we impose the slow-roll approximation.
	This requires
	$\epsilon = - \dot{H}/H^2 \ll 1$ where $H = \dot{a}/a$ is the
	Hubble parameter.
	We introduce an $\epsilon$-parameter
	for each species of light field,
	\begin{equation}
		\epsilon_\alpha \equiv \frac{1}{2\Mp^2}
			\frac{\dot{\phi}_\alpha^2}{H^2} ,
		\label{eq:epsilon-def}
	\end{equation}
	in terms of which
	one can write
	$\epsilon = \sum_\alpha \epsilon_\alpha$.
	The slow-roll approximation therefore entails $\epsilon_\alpha \ll 1$.
	
	\para{Huygens' equation}%
	Combining~\eqref{eq:epsilon-def} and the field equation for
	$\phi_\alpha$, and making use of the slow-roll
	approximation, we find
	\begin{equation}
		\frac{\d \phi_\alpha}{\d N}
		= \pm \Mp \sqrt{2 \epsilon_\alpha} 
		= - \Mp^2 \partial_\alpha \ln V ,
		\label{eq:field-eq}
	\end{equation}
	where $\d N \equiv \d \ln a$
	measures the number of e-foldings of expansion experienced
	along the trajectory,
	and $\partial_\alpha$ denotes a partial derivative with respect
	to $\phi_\alpha$.
	Eq.~\eqref{eq:field-eq} constrains
	the canonical momenta $\sim \dot{\phi}_\alpha$
	to lie on a submanifold of the classical phase space
	coordinatized by the fields $\phi_\alpha$.
	This simplification is a consequence of the slow-roll approximation.
	In a theory with $M$ scalar fields, it implies that we may
	work with the simpler $M$-dimensional field space
	instead of the full
	$2M$-dimensional phase space.
	This is convenient, although when
	we later
	abandon the slow-roll approximation
	we will have to
	return to phase space.

	In what follows we often rewrite~\eqref{eq:field-eq}
	in the form
	\begin{equation}
		\frac{\d \phi_\alpha}{\d N}
		= u_\alpha ,
		\;\;
		\mbox{where}
		\;\;
		u_\alpha \equiv - \Mp^2 \partial_\alpha f
		\;\;
		\mbox{and}
		\;\;
		f \equiv \ln \frac{V}{V_\ast} ,
		\label{eq:flow-equation}
	\end{equation}
	and interpret the solution $\phi_\alpha(N)$
	as an integral curve of the vector field $u_\alpha$,
	parametrized by $N$.
	The scale $V_\ast$ is arbitrary.
	Since $f$ is a gradient, these integral curves correspond to
	pure potential flow.%
		\footnote{Since $\d f / \d N = u_\alpha \partial_\alpha f
		= - \| u_\alpha \|^2 / \Mp^2$,
		it follows that
		$f$ is monotone decreasing along each integral curve.
		Therefore one may loosely think of $f$ as a Lyapunov function
		(or Morse function) for the flow.}
	
	The unit vector parallel to $u_\alpha$ is
	\begin{equation}
		\hat{n}_\alpha \equiv \frac{u_\alpha}{\Mp \refractiveindex} ,
		\label{eq:unit-vector}
	\end{equation}
	where we have defined 
	\begin{equation}
	\refractiveindex \equiv \sqrt{2\epsilon} .
	\end{equation}
	It follows that the arc length
	along an integral curve, labelled $s$
	and measured using a flat Euclidean metric
	on field space,
	satisfies $\d s = \Mp \refractiveindex \, \d N$.
	Reparametrizing each curve in terms of $s$, the flow
	equation~\eqref{eq:flow-equation} can be rewritten
	\begin{equation}
		\refractiveindex \frac{\d \phi_\alpha}{\d s}
			= - \Mp \partial_\alpha \ln V
			= \partial_\alpha S ,
		\label{eq:huygens}
	\end{equation}
	where $S \equiv - \Mp f$ is Hamilton's characteristic
	function.
	Eq.~\eqref{eq:huygens} is \emph{Huygens' equation}.
	Under the assumptions of geometrical optics,
	it describes the propagation
	of a light ray in a medium of spatially varying refractive index
	$\refractiveindex$.

	\para{Snell's law}%
	We conclude that the inflationary trajectories in field space are
	precisely the light rays of geometrical optics,
	for which
	Huygens' equation can
	be thought of as a generalization
	of the Ibn Sahl or Snell--Descartes law.
	The wavefronts correspond to level sets of the characteristic
	function $S$
	and are therefore equipotentials, or surfaces of constant energy
	density in field space.
	Each light ray is locally orthogonal to these surfaces,
	so the vector $\hat{n}_\alpha$ is locally the unit vector normal
	to a surface of constant energy density.

	When slow-roll is a good approximation $\nu$ is small,
	$\nu \ll 1$,
	and increases to $\nu \sim \Or(1)$
	near the end of inflation.
	
	\subsection{Rays on phase space}

	In some circumstances the slow-roll approximation
	is not available.
	This may be the case during inflation
	if slow-roll is temporarily violated---%
	perhaps during a turn in field space, to be studied
	in~\S\ref{sec:local-fnl}---%
	or on approach to the end of inflation,
	where $\epsilon \sim 1$.
	
	In such cases we must return to the full second-order field equation,
	which
	cannot be written in the form of Eq.~\eqref{eq:field-eq}.
	To obtain an analogue of geometrical optics
	one must pass to a Hamiltonian formalism.
	We define
	\begin{equation}
		p_\alpha = \frac{\d \phi_\alpha}{\d N} .
		\label{eq:momentum-def}
	\end{equation}
	This plays the role of Huygens' equation for $\phi_\alpha$.
	In terms of $p_\alpha$, the scalar field equation becomes
	\begin{equation}
		\frac{\d p_\alpha}{\d N}
		=
		[\epsilon(p) - 3] p_\alpha
		- \frac{V_\alpha(\phi)}{H(\phi,p)^2} .
		\label{eq:momentum-evolution}
	\end{equation}
	We must also rewrite $\epsilon$ and $H$ in terms of
	$p_\alpha$, obtaining
	\begin{subequations}
	\begin{align}
		\epsilon(p) \equiv
			- \frac{\dot{H}}{H^2} & = \frac{p_\alpha p_\alpha}{2 \Mp^2}
		\label{eq:non-sr-epsilon}
		\\
		H(\phi,p)^2 \Mp^2 & = \frac{V(\phi)}{3-\epsilon(p)} .
		\label{eq:non-sr-h}
	\end{align}
	\end{subequations}
	Note that $\epsilon$ is purely a function of $p_\alpha$,
	whereas $H$ is a function of both $\phi_\alpha$ and $p_\alpha$.
	
	Eqs.~\eqref{eq:momentum-def}--\eqref{eq:momentum-evolution}
	show that, beyond slow-roll, the \emph{precise}
	analogy with Huygens' equation is lost.
	Although these equations
	define a congruence of rays in phase space,
	it is not possible to find a characteristic function $S$
	so that these rays are everywhere orthogonal to equipotentials of $S$.
	Such a function would have to satisfy
	$\partial_{\phi_\alpha} S = p_\alpha$,
	and therefore $S = p_\alpha \phi_\alpha + g(p)$
	for arbitrary $g$.
	Unfortunately, there is no choice for $g$ which reproduces
	the right-hand side of Eq.~\eqref{eq:momentum-evolution}.
	
	The majority of our analysis requires only the first-order
	evolution equations~\eqref{eq:momentum-def}--\eqref{eq:momentum-evolution},
	and at this level the formalism we develop will apply
	to evolution in phase space without imposing slow-roll.
	For that purpose it is convenient to combine
	$\phi_\alpha$ and $p_\alpha$
	into a single phase-space coordinate. We continue
	to write this $\phi_\alpha$, with the understanding that
	$\alpha$ now ranges over the $2M$ dimensions of phase space.
	The velocity vector is likewise $u_\alpha$.

	\subsection{Jacobi fields and beam cross-sections}
	\label{sec:jacobi}
	
	To proceed, we must carry the initial distribution of
	occupation probabilities along the flow, forming the ``image''
	distribution of interest.
	In optical language, our task
	is to understand how images generated from
	a source of light rays are distorted by passage through a medium.
	
	It was explained above that the typical spacing between
	arbitrarily selected members of the ensemble should be
	roughly of order the
	quantum scatter, $\sigma \sim \langle \delta \phi^2 \rangle^{1/2}$.
	Because
	$\sigma/\Mp \sim 10^{-5} \ll 1$,
	this is small in comparison with the natural scale $\Mp$.
	Therefore
	the orbits traversed by the cloud
	trace out a narrowly-collimated spray or ``bundle'' of light rays
	in phase space.
	In canonical models of inflation, setting initial conditions
	near horizon-crossing will make
	the initial profile close to Gaussian
	\cite{Seery:2005gb}.
	Therefore the evolution of the ensemble is similar to
	the evolution of tightly-focused Gaussian laser beam propagating in an
	optical cavity.
		
	\para{Connecting vectors}%
	Cross-sections within the laser beam may be focused, sheared
	or rotated by
	refraction. These possibilities are
	familiar from the study of weak gravitational lensing.

	To obtain a quantitative description we slice the laser
	beam open, generating a cross-section.
	The precise slicing is arbitrary.
	For applications to inflation we will often
	slice along surfaces of fixed
	energy density, or after a fixed number of e-folds.
	Distortions of the cross section can be studied
	if we know how an arbitrary basis is transported from slice
	to slice.
	In general relativity this would be Fermi-Walker transport
	\cite{Hawking:1973uf}.
	
	Jacobi used this method to study geodesic deviation
	on Riemannian manifolds. For this reason
	an infinitesimal vector propagated along the beam
	is called a
	\emph{Jacobi field}.
	Taking $\delta \phi_\alpha$ to be such a field
	and the flow vector $u_\alpha$ to be sufficiently smooth,
	it will be transported by the equation
	\begin{equation}
		\frac{\d \delta \phi_\alpha}{\d N}
		 	= \delta \phi_\beta \partial_\beta u_\alpha
			= u_{\alpha \beta} \delta \phi_\beta .
		\label{eq:deviation-equation}
	\end{equation}
	The quantity $u_{\alpha \beta} \equiv \partial_\beta u_\alpha$
	is the \emph{expansion tensor}.
	It can be expanded in terms of a dilation $\theta = \tr u_{\alpha \beta}$,
	a traceless symmetric shear $\sigma_{\alpha \beta}$
	and an antisymmetric twist $\omega_{\alpha \beta}$,
	\begin{equation}
		u_{\alpha \beta} \equiv \frac{\theta}{d} \delta_{\alpha \beta}
			+ \sigma_{\alpha \beta} + \omega_{\alpha\beta} ,
		\label{eq:expansion-tensor}
	\end{equation}
	where $d = M$
	for flows on field space,
	or $d = 2M$ if we do not impose the slow-roll approximation and
	work on the full phase space.
	In either case $\delta_{\alpha\beta}$
	is the Kronecker $\delta$.
	
	\para{Optical scalars}%
	Dilation describes rigid,
	isotropic rescaling of $\delta \phi_\alpha$ by $1+\theta$.
	It represents
	a global tendency of the light rays to focus or defocus.
	The shear $\sigma_{\alpha\beta}$
	is a symmetric square matrix and can therefore
	be diagonalized, yielding $d$ eigenvalues $\xi_{\isoindex{i}}$
	and corresponding eigenvectors
	$s_{\alpha,\isoindex{i}}$ representing the principal shear
	directions (here $\isoindex{i}$ is a label
	taking values $1, \ldots, d$;
	see \S\ref{sec:local-fnl}).
	The shear describes a rescaling of the component of the
	connecting vector in the direction $s_{\alpha,\isoindex{i}}$ by
	a factor $1 + \xi_{\isoindex{i}}$.
	Tracelessness of $\sigma_{\alpha\beta}$ implies
	$\sum_{\isoindex{i}} \xi_{\isoindex{i}} = 0$,
	so expansion in one direction must be accompanied by contraction in
	another. Therefore shear preserves cross-sectional area.
	Finally, the twist $\omega_{\alpha\beta}$ describes a rigid
	volume-preserving rotation of $\delta \phi_\alpha$, representing a
	tendency of neighbouring trajectories to rotate around each other.
	
	It is useful to define $\sigma^2$ to satisfy
	\begin{equation}
		\sigma^2 \equiv \frac{1}{2} \sigma_{\alpha \beta} \sigma_{\alpha \beta}
		.
	\end{equation}
	Imposing the slow-roll approximation and working on field space,
	the flow is orthogonal to equipotentials of Hamilton's
	characteristic function.
	Therefore it is a pure
	potential flow,
	for which $\omega_{\alpha\beta} = 0$.
	On the full phase space this property is lost and
	the twist can be non-zero.
	In such cases
	it is helpful to define $2 \omega^2 = \omega_{\alpha\beta}
	\omega_{\alpha\beta}$. 
	Together, $\theta$, $\sigma^2$ and $\omega^2$
	comprise
	the optical scalars introduced by Sachs
	and Penrose
	\cite{Sachs:1961zz,*Penrose:1966}.
	
	\para{van Vleck matrix}%
	Eq.~\eqref{eq:deviation-equation}
	has a well-known formal solution
	in terms of an
	ordered exponential \cite{fried2002green}.
	This method was used
	Rigopoulos, Shellard \&
	van Tent~\cite{Rigopoulos:2005ae,*Rigopoulos:2005us},
	and later by
	Yokoyama {\etal} \cite{Yokoyama:2007uu}.
	It yields an explicit (but formal) expression for transport
	of any Jacobi field along the beam,
	\begin{equation}
		\delta \phi_\alpha(N)
			= \Gamma_{\alpha \beta}(N,N_0) \delta \phi_{\beta}(N_0) ,
		\label{eq:jacobi-transport}
	\end{equation}
	where $\delta \phi_{\beta}(N_0)$ is the Jacobi field on some initial slice
	$N = N_0$. Eq.~\eqref{eq:jacobi-transport} describes the evolution
	of this Jacobi field at any later time $N$.
	The matrix $\Gamma_{\alpha\beta}(N,N_0)$ satisfies
	\begin{equation}
		\Gamma_{\alpha\beta}(N,N_0) \equiv
			\pathorder \exp \int_{N_0}^N u_{\alpha \beta}(\varN) \, \d\varN
			,
		\label{eq:propagator}
	\end{equation}
	where the path-ordering operator $\pathorder$
	rewrites its argument with early times on the right-hand side,
	and later times on the left.
	We will occasionally refer to $\Gamma_{\alpha\beta}$ as the
	propagator matrix.
	It is closely related to a Wilson line.
	
	\label{para:index-convention}
	\subpara{Index convention}%
	Eq.~\eqref{eq:propagator} can be simplified
	with the aid of an index convention.
	Up to this point we have been labelling field-space indices
	using Greek symbols $\alpha, \beta$, etc.
	To avoid writing the time of evaluation explicitly,
	we adopt the convention that Greek indices denote evaluation at the
	late time of interest, $N$.
	Latin indices $i$, $j$, etc., denote evaluation at the early time
	$N_0$.
	Therefore $\Gamma$ can be written
	as a mixed index object, $\Gamma_{\alpha i}$.
	
	Eq.~\eqref{eq:jacobi-transport} immediately implies
	\begin{equation}
		\Gamma_{\alpha i}
			= \frac{\partial \phi_\alpha}{\partial \phi_i} ,
		\label{eq:gamma-deriv}
	\end{equation}
	and endows this derivative with a geometric interpretation.
	It plays
	an important role in the
	Lyth--Rodr\'{\i}guez
	implementation of the separate universe approximation \cite{Lyth:2005fi},
	where it appears due to a Taylor expansion
	in the initial conditions
	local to each $L$-sized patch.
	In this formulation, one often projects
	on to
	equipotential surfaces in field space.
	We define
	$h_{\alpha\beta} = \partial \phi^c_\alpha / \partial \phi_\beta$
	to obtain
	\begin{equation}
		\frac{\partial \phi^c_\alpha}{\partial \phi_i}
		=
		h_{\alpha\beta} \Gamma_{\beta i} .
		\label{eq:gamma-c-deriv}
	\end{equation}
	The notation `$c$' indicates that $\d \phi^c_\alpha$
	can be thought of as
	the variation of a field $\phi^c_\alpha$
	defined on a fixed comoving \emph{spacetime}
	hypersurface
	\cite{Vernizzi:2006ve,Battefeld:2006sz,*Seery:2006js}.
	It follows from
	geometrical aguments that $h_{\alpha\beta} = \delta_{\alpha\beta}
	- \hat{n}_\alpha \hat{n}_\beta$, where
	$\hat{n}_\alpha$ is	
	the unit normal to
	phase-space slices of constant potential energy,
	defined in~\eqref{eq:unit-vector}.
	The tensor
	$h_{\alpha\beta}$
	is the induced metric
	(or ``first fundamental form'')
	on these surfaces.
	Eq.~\eqref{eq:gamma-c-deriv} shows that choice of gauge is
	associated with projection onto an appropriate hypersurface
	in phase space.
	Moreover,
	Eqs.~\eqref{eq:gamma-deriv}--\eqref{eq:gamma-c-deriv}
	show that
	partial derivatives with respect to
	$\phi_i$ are associated with propagation of
	Jacobi fields along the bundle.
	
	\subpara{Caustics}%
	The matrix $\Gamma_{\alpha i}$ appears whenever it is necessary
	to track the distortion of a line element along a flow,
	and has applications in fluid dynamics, general relativity
	and elsewhere
	\cite{Hawking:1973uf,Visser:1992pz}.
	DeWitt--Morette observed that,
	considered as a matrix of Jacobi fields,
	Eq.~\eqref{eq:gamma-deriv}
	was related to the inverse of the van Vleck matrix,
	introduced in the construction of semiclassical
	(``WKB'') approximations to the path integral
	\cite{DeWittMorette:1976up,*DeWittMorette:1984du,*DeWittMorette:1984dw}.%
		\footnote{In DeWitt--Morette~\cite{DeWittMorette:1976up}
		the proof is ascribed to B.S. DeWitt.
		DeWitt--Morette noted that the relation between
		Jacobi fields and variation of a general solution of the
		field equations with respect to its constants of
		integration had been known to Jacobi
		(ultimately leading to his development of what is now
		Hamilton--Jacobi theory),
		and suggested that this technique could be used to simplify
		the long calculations which arise when solving
		Jacobi's equation.
		Applied to inflationary correlation functions, the history
		has been reversed: the
		variational formulae came first,
		in the form of the Lyth--Rodr\'{\i}guez algorithm.
		This
		often leads to simple analytic results,
		as DeWitt--Morette foresaw.
		But,
		as we explain in~\S\ref{sec:flow-deltaN},
		this method is unsuited to numerical implementation,
		because of the small numerical tolerances required
		to reliably determine variation with respect to the initial conditions.
		It is preferable to solve an \emph{ordinary}
		differential equation, such as Jacobi's
		equation~\eqref{eq:deviation-equation} or~\eqref{eq:jacobi}.}
	We define
	\begin{equation}
		\Gamma^{-1}_{i \alpha} =
		\delta_{i \beta}
		\pathorder
		\exp
		\left(
			- \int_{N_0}^{N} u_{\beta \alpha}(\varN) \; \d\varN
		\right) .
	\end{equation}
	The van Vleck matrix is
	$\Delta_{i \alpha}
	\equiv (N-N_0)^d \Gamma^{-1}_{i \alpha}$, and has a well-known
	interpretation in geometrical optics as a measure of focusing
	or defocusing:
	in particular, $|\det \Delta| \rightarrow \infty$
	at a
	\emph{caustic}, where light rays converge.
	Since $(N-N_0)$ is nonzero for $N \neq N_0$,
	a singularity in the van Vleck determinant
	implies a singularity in $\det \Gamma^{-1}$.
	Applying~\eqref{eq:expansion-tensor}, we conclude
	\begin{equation}
		\frac{1}{\det \Gamma^{-1}}
			= \det \Gamma
			\equiv \Theta(N,N_0)
			= \exp \int_{N_0}^N \theta(\varN) \; \d\varN .
		\label{eq:focusing-def}
	\end{equation}
	Therefore $\Theta \rightarrow 0$ at a caustic.
	This happens after finitely many e-folds only if
	$\theta \rightarrow -\infty$
	during the flow.
	Otherwise, $\Theta$ is decreasing in regions where
	$\theta$ is negative, with large negative $\theta$
	implying strong focusing. Large positive $\theta$
	implies strong defocusing.
	More generally the propagator matrix
	can be rewritten in terms of $\Theta$, giving
	\begin{equation}
		\Gamma_{\alpha i}
			= \Theta(N,N_0)^{1/M}
			\pathorder
			\exp
			\left(
				\int_{N_0}^N
				(\sigma + \omega)_{\alpha\beta}(\varN)
				\; \d\varN
			\right)
			\delta_{\beta i} .
		\label{eq:gamma-decomposition}
	\end{equation}
	The ordered exponential has determinant unity
	and therefore does not change the cross-sectional area of the bundle.

	\subsection{Adiabatic limit}
	\label{sec:adiabatic}

	Caustics
	have an important interpretation in the flows
	describing an inflationary model.
	If the bundle of trajectories has finite cross
	section, then the ensemble contains members which are evolving
	along multiple phase space trajectories.
	These are the eponymous ``separate universes''
	with their individual initial conditions.
	
	Under these circumstances one or more isocurvature modes
	exist.
	These are connecting vectors which relate the different
	$\phi_\alpha$
	within the bundle which all lie on a surface of fixed energy density,
	say $\Sigma_{\rho}$.
	Their number is determined
	by the rank of $h_{\alpha \beta} \Gamma_{\beta i}$.
	In
	the special case where the bundle cross-section decays to a point,
	there is a unique
	intersection between the bundle and $\Sigma_\rho$.
	Therefore
	$h_{\alpha \beta} \Gamma_{\beta i}$
	has rank zero
	and all isocurvature modes disappear.
	In this limit, each member of the ensemble traverses the
	same orbit,
	differing from the others
	only by its relative position,
	which corresponds to the adiabatic mode,
	$\zeta$.
	It follows that,
	when the cross-section collapses to a point,
	the fluctuations become purely adiabatic.
	Elliston~{\etal}~\cite{Elliston:2011dr}
	described this as an `adiabatic limit'.
	After this limit has been reached $\zeta$ is conserved
	\cite{Lyth:2004gb,Rigopoulos:2003ak}.

	Flows which reach an adiabatic limit
	during inflation
	are no more or less likely---%
	or natural---from the viewpoint of fundamental physics.
	But flows reaching an adiabatic limit \emph{are} more predictive,
	because a perturbation in the purely adiabatic mode
	remains adiabatic long after inflation ends
	\cite{Weinberg:2008nf,*Weinberg:2008si},
	even during epochs for which we are ignorant of the
	relevant physics.
	Contrariwise,
	if any isocurvature modes remain
	then
	members of the ensemble may
	rearrange their relative positions
	until these modes decay.
	This possibility was emphasized by
	Meyers \& Sivanandam \cite{Meyers:2010rg,*Meyers:2011mm};
	see also Ref.~\cite{Elliston:2011dr}.
	If the flow does not reach an adiabatic limit during inflation
	then the model is not predictive until we supply a prescription for
	the post-inflationary era,
	and
	observational predictions can depend on this choice.

	\para{Trivial, adiabatic and nonadiabatic caustics}%
	The outcome of this discussion is that
	approach to an adiabatic limit
	can be associated with convergence to a caustic.
	An early discussion of this principle,
	phrased almost precisely in these terms, was given by
	Wands \& Garc\'{\i}a-Bellido~\cite{Wands:1996kb}.
	We conclude that
	$\Theta \rightarrow 0$ is a necessary condition for
	an adiabatic limit to occur, but as we now explain it is not sufficient.%
		\footnote{One
		may have some reservations about this conclusion,
		because it seems to violate
		the Liouville theorem which guarantees
		conservation of phase-space volume.
		However, it should be remembered that the \emph{canonical}
		phase space coordinate to which Liouville's theorem applies
		are not the field-space position and momenta which we are
		using. In particular, the canonical momenta will typically include
		powers of the scale factor $a$.}
	A caustic can be classified by the number of dimensions lost by the
	flow, or equivalently the number of null eigenvalues
	of the propagator $\Gamma_{\alpha i}$
	at the caustic.
	An adiabatic limit is the special case where
	$\Gamma_{\alpha i}$ retains a \emph{single} non-null eigenvalue,
	but $h_{\alpha \beta} \Gamma_{\beta i}$ has
	\emph{no} non-null eigenvalues.
	We describe caustics which satisfy this condition as
	\emph{adiabatic}.
	
	Eq.~\eqref{eq:gamma-decomposition} shows that, were
	the integrated
	shear and twist to remain bounded while $\Theta \rightarrow 0$,
	then $\Gamma_{\alpha i} \rightarrow 0$. In this case no perturbations
	would survive,
	and we describe the caustic as \emph{trivial}.
	An example is the case where $u_{\alpha\beta}$ is pure dilation.
	But barring an accurate cancellation of this kind,
	at least some component of $(\sigma + \omega)_{\alpha\beta}$
	will typically scale proportionally
	to $\theta$
	on approach to the caustic.%
		\footnote{In principle
		$u_{\alpha\beta}$ could contain off-diagonal
		terms which grow \emph{faster} than the diagonal terms,
		and therefore $\theta$.
		In this case there could be a subspace of growing
		perturbations.
		If the growth is exponential this usually signals
		an instability, and the formalism we are describing becomes
		invalid.}

	\subpara{Shear opposes focusing}%
	If the perturbations are not to vanish completely, then some
	\emph{anisotropic}
	effect of shear and twist must oppose the isotropic contraction
	due to $\Theta \rightarrow 0$.
	
	First suppose the twist is negligible.
	We assume that the eigenvectors of $\sigma$
	stabilize in the vicinity of the caustic.
	If the shear has some number of positive eigenvalues
	$\lambda_i$ for which $\lambda_i/\theta$ has a finite, nonzero
	limit, then perturbations may survive in the subspace spanned
	by their corresponding eigenvalues.
	Tracelessless of $\sigma$ implies that
	at least one eigenvalue must be negative,
	and
	perturbations in the subspace spanned by the corresponding
	eigenvectors will disappear. Hence, at least one dimension will
	be lost by the flow.
	In practice it is often simpler
	to work directly with the
	eigenvalues of the
	expansion tensor $u_{\alpha\beta}$.
	
	If more than one eigenvalue of $\sigma$ is positive,
	then perturbations may survive in a two- or higher dimensional
	subspace.
	In this case the caustic does not describe approach to an
	adiabatic limit,
	and we call it \emph{nonadiabatic}.
	To obtain predictions for observable quantities
	the evolution must be continued.
	In practice this would require introduction of a
	reduced phase space describing only the surviving perturbations.
	The flow can then be followed in this reduced phase space
	until a further focusing event occurs.
	This may itself be an
	adiabatic limit, or might simply describe further reduction
	in the phase space.
	One should continue in this way until an adiabatic limit
	is finally achieved.
	An example of this behaviour could occur soon after the onset
	of slow-roll inflation.
	In the early stages, independent fluctuations in the
	field velocities survive. But
	when slow-roll is a good approximation these will be exponentially
	suppressed, making $\Theta$ become very small.
	One should therefore replace the full description by
	a reduced phase space which includes only field perturbations.
	In doing so one arrives at the field-space
	description of slow-roll inflation given in \S\ref{sec:slow-roll}.
	
	\subpara{Twist opposes focusing}%
	In slow-roll inflation, which we discuss in
	\S\ref{sec:slow-roll-focusing} below,
	a diverging shear
	is the only mechanism by which perturbations can survive
	on approach to a caustic.
	Where the twist is non-zero, which occurs when we do not
	impose the slow-roll approximation, more possibilities
	exist.
	Ultimately
	these must be addressed,
	to describe approach to an adiabatic limit
	when slow-roll is no longer a good approximation, but
	we defer this discussion for future work.
	
	\subsection{Focusing in the slow-roll approximation}
	\label{sec:slow-roll-focusing}

	In this subsection we give a more detailed discussion of the
	approach to a caustic during an era of slow-roll
	inflation.
	
	\para{Raychaudhuri equations}%
	Parametrizing each trajectory by e-folding number $N$,
	Eq.~\eqref{eq:flow-equation}
	constitutes an autonomous dynamical system.
	Therefore
	a derivative along the flow can be written
	$\d/\d N = u_\alpha \partial_\alpha$.
	In the absence of a nontrivial field-space metric
	all derivatives commute,
	and therefore
	$[\partial_\gamma, \partial_\beta] u_\alpha
	= 0$.
	Contracting with $u_\gamma$ and rearranging terms, one finds
	\begin{equation}
		\frac{\d u_{\alpha \beta}}{\d N}
			= \partial_\beta a_\alpha - u_{\alpha \gamma} u_{\gamma \beta} ,
	\end{equation}
	where $a_\alpha$ is the acceleration vector, defined by
	$a_\alpha = \d u_\alpha / \d N = u_\beta \partial_\beta u_\alpha$.
	For a potential flow, this can be simplified;
	comparison with Eq.~\eqref{eq:flow-equation} shows that
	\begin{equation}
		a_\alpha = \frac{\Mp^2}{2} \partial_\alpha \refractiveindex^2 ,
	\end{equation}
	where, as above, $\refractiveindex$ is the local refractive index.
	
	The evolution equations for the dilation and shear can be written
	\begin{widetext}
	\begin{subequations}
	\begin{align}
		\label{eq:raychaudhuri-dilation}
		\frac{\d \theta}{\d N} & =
			\Mp^2 \Hessian- \frac{\theta^2}{M} - 2 \sigma^2 \\
		\label{eq:raychaudhuri-shear}
		\frac{\d \sigma_{\alpha\beta}}{\d N} & =
			\Mp^2 \Big(
				\Hessian_{\alpha\beta} - \frac{\Hessian}{M} \delta_{\alpha\beta}
			\Big)
			- \frac{2 \theta}{M} \sigma_{\alpha\beta}
			- \Big(
				\sigma_{\alpha \gamma} \sigma_{\gamma \beta}
				- \frac{2 \sigma^2}{M} \delta_{\alpha\beta}
			\Big) .
	\end{align}
	\end{subequations}
	\end{widetext}
	These are commonly known as the Raychaudhuri equations.
	An equation for the evolution of
	the twist could be found in the same way,
	but is not needed in the slow-roll approximation.
	
	We have defined $\Hessian_{\alpha\beta}$ to be the Hessian of
	$\refractiveindex^2$,
	\begin{equation}
		\Hessian_{\alpha\beta} \equiv
			\frac{1}{2} \partial_\alpha \partial_\beta
			\refractiveindex^2 ,
	\end{equation}
	and $\Hessian$ is its trace.
	Because the Hessian measures the local curvature of a function,
	one can regard $\Hessian_{\alpha\beta}$ as a
	measure of the curvature of surfaces of constant
	refractive index in field space.
	
	\para{Jacobi equation}%
	Eq.~\eqref{eq:deviation-equation}
	shows that Jacobi fields oriented along eigenvectors
	of $u_{\alpha\beta}$ with positive eigenvalues grow, whereas
	those oriented along eigenvectors with negative eigenvalues decay.
	
	We can find an alternative description in terms of the
	refractive index $\refractiveindex$.
	Taking a derivative of~\eqref{eq:deviation-equation}
	along the flow and
	using the Raychaudhuri equations
	to eliminate derivatives of the dilation and shear
	yields the Jacobi equation,%
		\footnote{When using Jacobi fields to study geodesic deviation on
		a Riemannian manifold, this equation takes the form
		$\delta
		\ddot{\phi}_\alpha = - R_{\alpha \hat{n} \beta \hat{n}}
		\delta \phi_\beta$,
		where $R_{\alpha \hat{n} \beta \hat{n}}
		= \hat{n}^\rho \hat{n}^\sigma R_{\alpha \rho \beta \sigma}$
		is a component of the Riemann curvature
		projected along the tangent to the geodesic.}
	\begin{equation}
		\frac{\d^2 \delta \phi_\alpha}{\d N^2} =
			\Mp^2 \Hessian_{\alpha\beta} \delta \phi_\beta .
		\label{eq:jacobi}
	\end{equation}
	It follows that the behaviour of the Jacobi fields is
	determined by the curvature of $\refractiveindex^2$,
	considered as a function in field space.
	(Note this is related to, but not the same as,
	the curvature of surfaces of
	constant $\refractiveindex$.)
	Qualitatively, Jacobi fields oriented along eigenvectors of
	$\Hessian_{\alpha\beta}$ with negative eigenvalues---directions
	of negative curvature---will have
	quasi-trigonometric solutions.
	These will pass through zero,
	corresponding to the collapse of some Jacobi fields
	to zero length.
	Fields oriented along eigenvectors with positive eigenvalues
	will have exponential solutions. Unless the initial conditions
	are precisely adjusted, these will typically grow.

	\para{Focusing theorem}%
	By adapting
	the geodesic focusing theorem of general relativity~\cite{Hawking:1973uf}
	we can determine the circumstances under which focusing will occur
	after finitely many e-folds.
	Pick a point on the flow where the expansion is negative,
	with value $\theta_\star < 0$.
	Inspection of~\eqref{eq:raychaudhuri-dilation}
	shows that, if $\Hessian < 0$,
	then $\theta \rightarrow -\infty$ within
	$\Delta N = M/|\theta_\star|$
	e-folds, where $M$ is the dimension
	of field space. 
	Any point where $\theta = -\infty$ is a caustic,
	because on arrival at this point
	$\Theta = 0$.
	
	Since
	Morse's lemma
	implies that $\Hessian$ is negative
	in a neighbourhood of any
	local maximum of the refractive index,
	$\refractiveindex^2 = 2\epsilon$,
	one might hope to associate such local maxima with terminal
	points for inflation
	at which an adiabatic limit would be nearly achieved.
	
	However, the conditions of the focusing theorem are not satisfied for
	typical potentials.
	More usually
	the slow-roll approximation
	forces all fields to settle into a terminal vacuum increasingly
	slowly, requiring an infinite number of e-folds to reach $\Theta = 0$.
	Moreover,
	in practical examples the slow-roll approximation
	will break down and inflation will terminate long before the
	caustic is reached.
	Therefore we should \emph{not} expect to achieve
	precisely
	$\Theta = 0$ during inflation.
	Nevertheless, a model may be
	sufficiently predictive if the flow
	spends enough e-folds in a
	region of large negative $\theta$
	that $\Theta$ is exponentially suppressed before
	inflation ends.
	
	In simple potentials it is often clear when $\zeta$
	ceases to evolve.
	But for more complicated potentials the situation may not be
	so clear.
	Within the slow-roll approximation,
	this discussion shows that
	$\Theta \gtrsim 1$ can be taken
	as a clear indication that isocurvature modes are still present.
	Their future decay is likely to influence $\zeta$ and the outcome of any
	calculation which terminates
	with $\Theta \gtrsim 1$ should not be considered a prediction for
	observable quantities.
	Conversely,
	$\Theta \ll 1$ is an indication that some
	decay of
	isocurvature modes has taken place.
	The precise nature of the decay must be deduced from the
	behaviour of the shear and twist.
	If perturbations survive only in a one-dimensional subspace
	than we can infer that the isocurvature modes have
	decayed to the point that
	$\zeta$ will be approximately conserved.

	\para{Example: quadratic Nflation}%
	We illustrate these ideas using
	the quadratic approximation to Nflation
	\cite{Dimopoulos:2005ac,Kim:2006ys,*Kim:2010ud,*Kim:2011jea}.
	The potential is
	\begin{equation}
		V = \sum_\alpha \frac{1}{2} m_\alpha^2 \phi_\alpha^2 .
	\end{equation}
	This model is of interest in its own right, but also describes the
	approach to a generic stable minimum after suitable choice of field
	space coordinates. We suppose that there is at least
	a modest hierarchy
	among the masses, and order these so that $m_\alpha < m_\beta$
	if $\alpha < \beta$.
	The most massive field will settle into its minimum first, followed by
	the next most massive field. Therefore approach to the final minimum
	will be described by a trajectory on which only $\phi_1$ is dynamical,
	with all other $\phi_\alpha$ approximately zero.
	We describe this as the ``inflow'' trajectory.
	
	On the inflow trajectory, the dilation satisfies
	\begin{equation}
		\theta^{\text{inf}} \approx
			- 2 \frac{\Mp^2}{\phi_1^2} \left(
				\sum_{\alpha \geq 2} \frac{m_\alpha^2}{m_1^2} - 1
			\right) .
	\end{equation}
	The minimum $\phi_1 = 0$ is a caustic,
	but
	as discussed above
	it cannot be reached
	after finitely many e-folds (within the slow-roll approximation).
	The expansion tensor
	satisfies
	\begin{equation}
		u_{\alpha\beta}^{\text{inf}}
		\approx
		\left(
			\begin{array}{cccc}
				\displaystyle 2 \frac{\Mp^2}{\phi_1^2}
				\\
				& \ddots
				\\
				& & \displaystyle - 2 \frac{m_\alpha^2}{m_1^2}
				\frac{\Mp^2}{\phi_1^2}
				\\
				& & & \ddots
			\end{array}
		\right) .
		\label{eq:inflow-expansion-tensor}
	\end{equation}
	This has one positive eigenvalue and the rest negative,
	so we expect it will correspond to an adiabatic limit.
	
	The $(1,1)$ component of
	$\Gamma^{\mathrm{inf}}$
	diverges near the caustic.
	This does not signal an instability, but only that
	$\delta \phi_1$
	grows at precisely the required rate to
	give constant $\zeta \sim (H/\dot{\phi}_1) \delta \phi_1$.
	
\begin{figure*}
	
		\small

		\vspace{7mm}
		\hfill
		\begin{fmfgraph*}(150,90)
			\fmfpen{thin}
			\fmfipair{a,b,c,d,e}
			\fmfiequ{a}{(w,h)}
			\fmfiequ{b}{(0.4w,0.9h)}
			\fmfiequ{c}{(0.1w,0.7h)}
			\fmfiequ{d}{(0,0.3h)}
			\fmfiequ{e}{(0,0)}
			\fmfi{fermion,label=$\Gamma^\ast$,label.side=right}{a .. b .. c .. d}
			\fmfi{dashes,label=$\Gamma^{\text{inf}}$,label.side=right}{d .. e}
			\fmfiv{decor.shape=circle,decor.filled=full,decor.size=0.025w,label={initial point},label.angle=0}{a}
			\fmfiv{label=$\phi_1^\ast$,label.angle=0}{d}
			\fmfiv{decor.shape=circle,decor.filled=full,decor.size=0.025w,label={focus point $\phi_1 = 0$},label.angle=0}{e}
		\end{fmfgraph*}
		\hfill
		\mbox{}

		\caption{Decomposition of propagator along an inflationary
		trajectory. Trajectories flowing into the minimum
		from most initial points join an ``inflow trajectory''
		(represented by a dashed line)
		at $\phi_1 = \phi_1^\ast$. The precise location
		of the junction is
		initial-condition dependent.
		The inflow trajectory sinks into the
		caustic, which here is a focus point,
		giving nearly-universal behaviour
		in the final stages of approach.
		This parallels the discussion of universality in critical
		phenomena; however, here, the universal region is often
		physically inaccessible because
		the slow-roll approximation breaks down in the vicinity
		of the focus point.
		The remaining part of the trajectory
		(represented by a solid line) is non-universal,
		and typically cannot be calculated analytically.\label{fig:inflow}}

	\end{figure*}
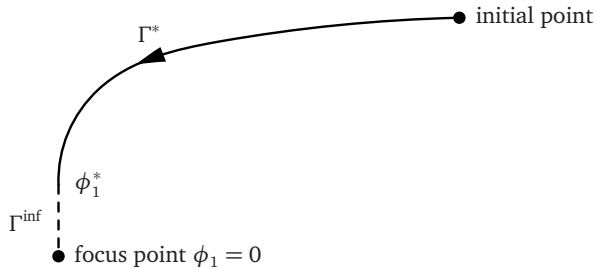

	Ordered exponentials
	such as~\eqref{eq:propagator} satisfy a composition property,
	allowing the integral over the inflationary trajectory
	to be broken in two.
	(See Fig.~\ref{fig:inflow}.)
	The first component is an integral from the initial point until
	the onset of the inflow trajectory.
	We take this to occur at $\phi_1 = \phi_1^\ast$,
	and choose $\phi_1^\ast$
	so that~\eqref{eq:inflow-expansion-tensor} is a good
	approximation there.
	The propagator at this point is
	$\Gamma^\ast_{\alpha i}$.
	It is a complicated
	weighted average over the
	trajectory, and cannot usually be calculated analytically.
	The second component is an integral over the inflow trajectory,
	which we denote $\Gamma^{\text{inf}}_{\alpha\beta}$.
	Therefore
	$\Gamma_{\alpha i} = (\Gamma^{\text{inf}}
	\Gamma^\ast)_{\alpha i}$.
	The inflow part can be
	computed from~\eqref{eq:inflow-expansion-tensor},
	\begin{equation}
		\Gamma^{\text{inf}}
		\approx
		\left(
			\begin{array}{cccc}
				\displaystyle \frac{\phi_1^\ast}{\phi_1} \\
				& \ddots \\
				& & \displaystyle \left( \frac{\phi_1}{\phi_1^\ast} \right)
					^{m_\alpha^2/m_1^2} \\
				& & & \ddots
			\end{array}
		\right) .
		\label{eq:inflow-propagator}
	\end{equation}
	Except perhaps for special choices of initial conditions,
	Eq.~\eqref{eq:inflow-propagator} gives rank $r=1$ at the caustic.
	Therefore this is an example of an adiabatic caustic.
	
	In more general circumstances, it may be necessary to diagonalize
	$u_{\alpha\beta}^\text{inf}$
	before integrating over the inflow trajectory.
	This is reminiscent
	of the introduction of
	scaling operators in a renormalization-group
	framework.
	Indeed,
	the entire analysis, and
	the emergence of rational but
	non-integer power-law scaling
	near the caustic,
	parallels a renormalization group analysis in the neighbourhood
	of a fixed point~\cite{Cardy:1996xt};
	compare also Eq.~(79) of Vernizzi \& Wands~\cite{Vernizzi:2006ve}.
	
\begin{figure*}
		\hfill
		\subfloat[][Dilation $\theta$\label{fig:dq-dilation}]{
			\includegraphics[scale=0.6]{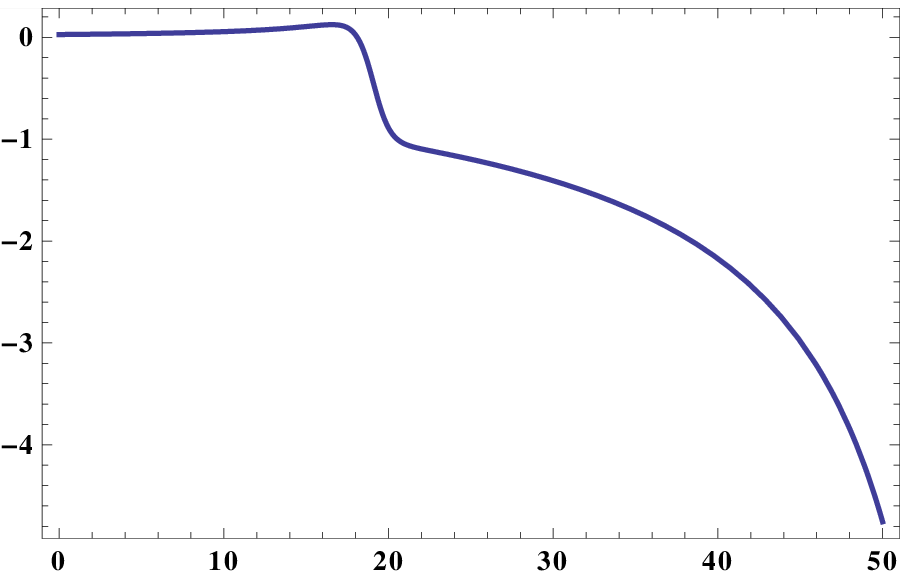}
		}
		\hfill
		\subfloat[][Integrated dilation $\int \theta \, \d N$
			\label{fig:dq-int-dilation}]{
			\includegraphics[scale=0.6]{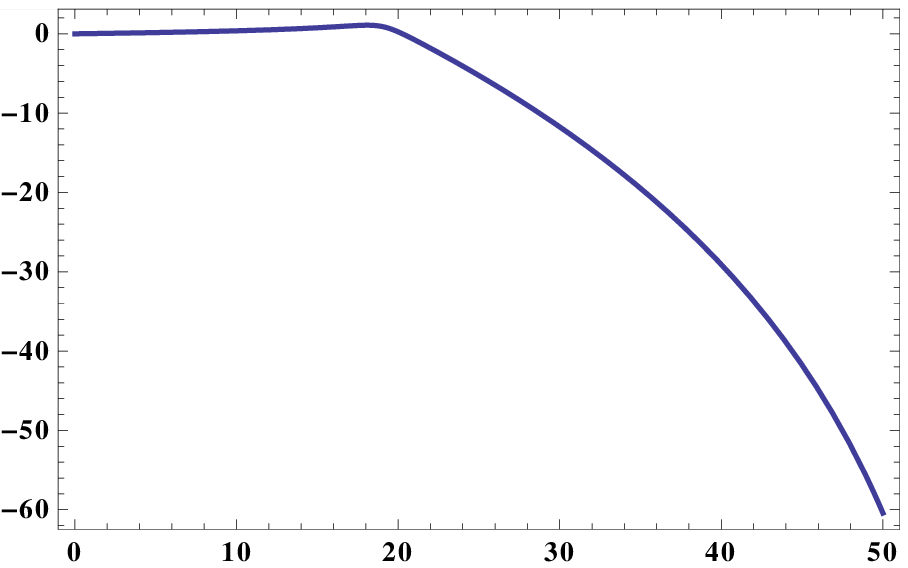}
		}
		\hfill
		\subfloat[][Focusing $\Theta = \e{\int \theta \, \d N}$
			\label{fig:dq-focusing}]{
			\includegraphics[scale=0.6]{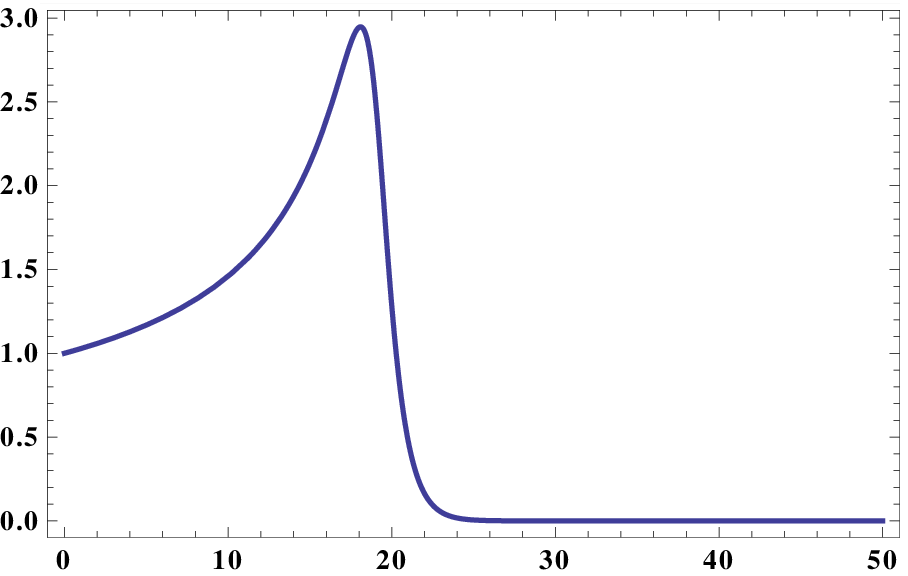}
		}
		\hfill
		\mbox{}
		\caption{Dilation, integrated dilation and focusing parameters
		in the double quadratic inflation model
		$V = \frac{1}{2} m_1^2 \phi_1^2 + \frac{1}{2} m_2^2 \phi_2^2$.
		The mass ratio is $m_1/m_2 = 9$,
		and the initial conditions are
		$\phi_1 = 8.2 \Mp$, $\phi_2 = 12.9 \Mp$.
		All plots are against the e-folding number $N$,
		measured from horizon exit of the mode in question.
		\label{fig:focusing-dq}}
	\end{figure*}

	\subpara{Focusing in double quadratic model}%
	Away from the inflow trajectory it is usually necessary to
	proceed numerically.
	In Fig.~\ref{fig:focusing-dq}
	we show the evolution of
	the focusing parameters
	in the well-studied model of double quadratic inflation
	\cite{Silk:1986vc,Polarski:1994rz,GarciaBellido:1995qq,Langlois:1999dw,
	Vernizzi:2006ve,Alabidi:2005qi,
	Rigopoulos:2005ae,*Rigopoulos:2005us}.
	The potential is $V = m_1^2 \phi_1^2 / 2 + m_2^2 \phi_2^2 / 2$.
	We choose the mass ratio $m_1/m_2 = 9$ and
	set initial conditions
	$\phi_1 = 8.2 \Mp$ and $\phi_2 = 12.9 \Mp$.

	Initially the evolution is mostly in the $\phi_2$ direction.
	When $\phi_2$ reaches the vicinity of its minimum there is a turn
	in field-space,
	which generates a spike in $\fNL$.
	After the turn,
	the inflow trajectory is reached along the $\phi_1$ direction.
	
	This evolution is reflected in the evolution of the bundle.
	Initially $\theta > 0$ and the cross-section slowly
	dilates. It reaches a maximum
	at roughly three times the original cross-sectional area.
	After the turn, $\theta$ rapidly drops to a negative value,
	and thereafter diverges exponentially to $-\infty$.
	Therefore the bundle-cross section very rapidly
	diminishes to almost zero cross-sectional area.
	This corresponds to an
	approximate caustic, and leads to an
	adiabatic limit.

	Eventually the divergence in $\theta$ would be cut off by
	a breakdown of the slow-roll approximation, but
	for typical parameter choices
	$\Theta$ will already be exponentially small
	at this point.
	
	\subpara{Example: axion-quadratic model}%
	Elliston~{\etal}~\cite{Elliston:2011dr}
	introduced an approximation to the hilltop
	region of axion N-flation \cite{Kim:2010ud, *Kim:2011jea}.
	The Hubble rate is dominantly supported by many axions
	in the quadratic region of their potential,
	and can be approximated by a single field.
	A few axions remain in the vicinity of the hilltop, where their
	contribution to $H$ is negligible but their contribution
	to the three- and higher $n$-point functions
	in the adiabatic limit is large.
	
\begin{figure*}
		\hfill
		\subfloat[][Dilation $\theta$\label{fig:aq-dilation}]{
			\includegraphics[scale=0.6]{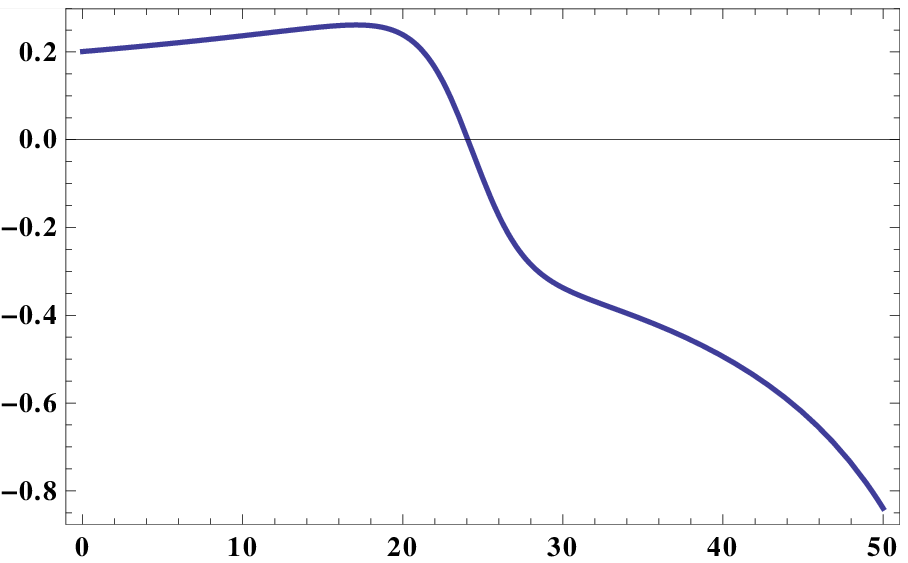}
		}
		\hfill
		\subfloat[][Integrated dilation $\int \theta \, \d N$
			\label{fig:aq-int-dilation}]{
			\includegraphics[scale=0.6]{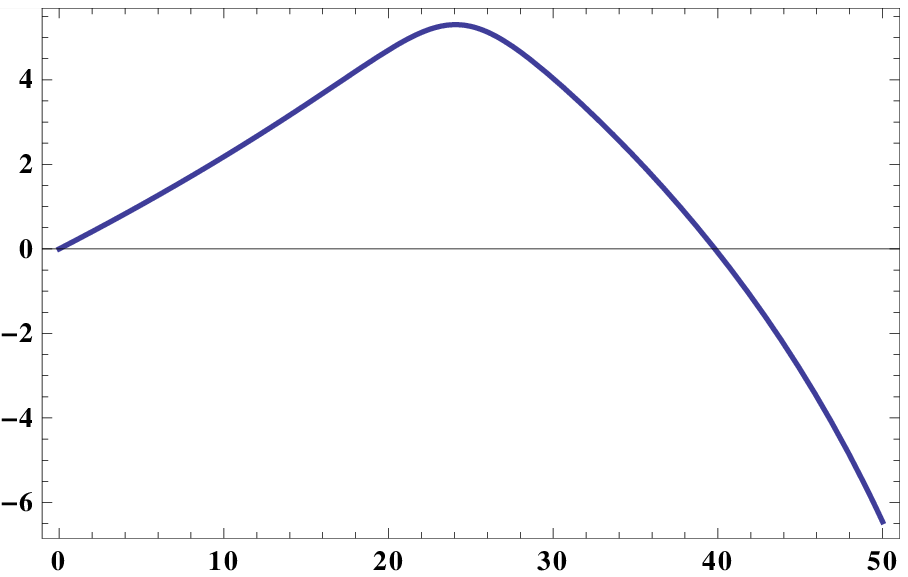}
		}
		\hfill
		\subfloat[][Focusing $\Theta = \e{\int \theta \, \d N}$
			\label{fig:aq-focusing}]{
			\includegraphics[scale=0.6]{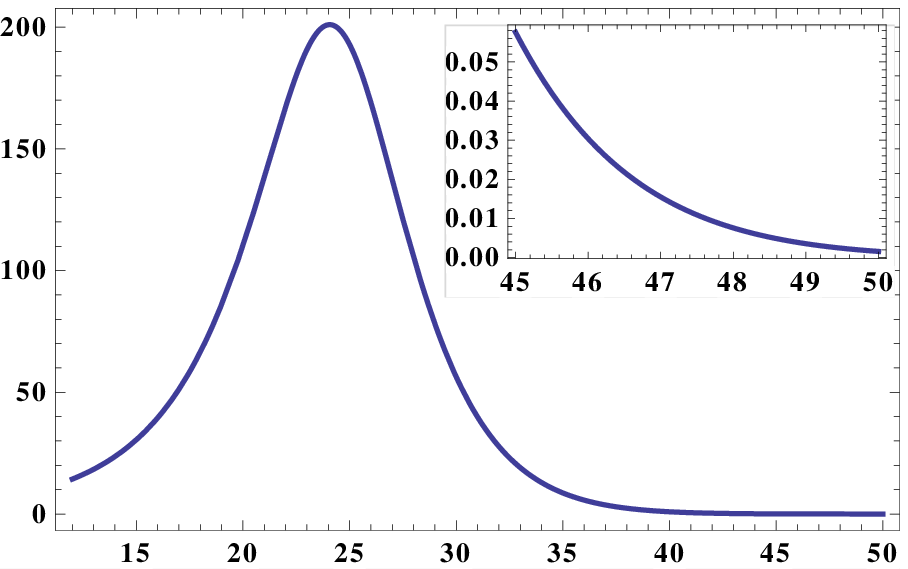}
		}
		\hfill
		\mbox{}
		\caption{Dilation, integrated dilation and focusing parameters
		in the axion-quadratic model
		$V = \frac{1}{2} m^2 \phi^2 + \Lambda^4 ( 1 - \cos 2\pi\chi /f )$.
		We have set $\Lambda^4 = 25 m^2 f^2 / 4\pi^2$
		and $f=\Mp$.
		The initial conditions are
		$\phi = 16 \Mp$, $\chi = (f/2 - 0.001) \Mp$.
		In (c), the inset panel shows the evolution
		of $\Theta$ near the end of inflation.
		All plots are against the e-folding number $N$,
		measured from horizon exit of the mode in question.
		\label{fig:focusing-aq}}
	\end{figure*}

	The potential is $V = m^2 \phi^2 / 2 + \Lambda^4(1-\cos 2\pi \chi/f)$.
	We set $\Lambda^4 = 25 m^2 f^2 / 4\pi^2$ and choose $f = \Mp$.
	In Fig.~\ref{fig:focusing-aq} we show the evolution
	for initial conditions $\phi = 16\Mp$
	and $\chi = (f/2-0.001) \Mp$.
	
	The evolution is similar to the double quadratic model.
	Initially $\theta$ is positive and the cross-sectional area grows.
	At its peak, it is more than 200 times the original
	cross-section.
	Eventually $\phi$ approaches its minimum and the Hubble friction
	decreases to the point that $\chi$ can evolve. It rolls away from the
	hilltop, eventually ending inflation.
	During this phase
	$\theta$ switches sign,
	ultimately diverging exponentially to $-\infty$.
	Therefore we approach an adiabatic limit.
	However,
	Fig.~\ref{fig:aq-focusing} shows that
	the rate of approach is quite slow.
	The cross-section decays softly,
	and by the end of inflation
	$\Theta \sim 10^{-3}$.
	Therefore an approximate adiabatic limit is reached
	and we can expect the observables to be roughly conserved
	through the post-inflationary evolution.

	\section{Transport equations}
	\label{sec:transport}
	
	We now apply these ideas to obtain evolution (or ``transport'')
	equations for the correlation functions in a fixed, comoving spacetime
	volume.
	In this section our analysis will be general,
	and can be applied to any perturbations whose evolution
	equations can be expressed in the form of Eq.~\eqref{eq:k-deviation}.
	If necessary this
	can be achieved
	as described in~\S\ref{sec:geo-optics}
	by passing to a Hamiltonian framework.
	It follows that the transport of correlation functions is most
	naturally expressed in phase space.
	
	\para{Connecting vectors}%
	Consider the set-up described
	in \S\ref{sec:introduction}, in which a comoving
	spacetime region of size $\coscale$ is smoothed into separate universes
	of size $L$.
	Pick any one of these $L$-sized regions, which we take to be
	at spatial position $\vect{x}$. The separate universe approximation asserts
	that the evolution of the smoothed fields in this region
	is given by the flow equation~\eqref{eq:flow-equation}.
	We denote the difference between these values and those in some other
	region, located at position
	$\vect{x} + \vect{r}$, by
	$\delta \phi_\alpha(\vect{r})$.
	This is a connecting vector in the sense of
	Eq.~\eqref{eq:deviation-equation}.
	Taylor expanding $u_\alpha$, the corresponding deviation equation is
	\begin{widetext}
	\begin{equation}
		\frac{\d \delta \phi_\alpha(\vect{r})}{\d N}
		= u_{\alpha\beta}[\phi(\vect{x})] \delta \phi_\beta(\vect{r})
		+ \frac{1}{2} u_{\alpha\beta\gamma}[\phi(\vect{x})]
			\left\{
				\delta \phi_\beta(\vect{r})
				\delta \phi_\gamma(\vect{r})
				-
				\langle
				\delta \phi_\beta(\vect{r})
				\delta \phi_\gamma(\vect{r})
				\rangle
			\right\}
		+ \cdots .
	\label{eq:real-deviation}
	\end{equation}
	We assume $\langle \delta \phi_\alpha(\vect{r}) \rangle = 0$
	and have subtracted a zero-mode to
	preserve this throughout the motion.%
		\footnote{In the language of Feynman
		diagrams, this would correspond to removing contributions
		arising from disconnected pieces.
		This procedure is routine in applications of the separate universe
		principle.}
	The tensor $u_{\alpha\beta}$
	was defined in~\eqref{eq:deviation-equation},
	and $u_{\alpha\beta\gamma} \equiv \partial_\gamma u_{\alpha\beta}$.
	We describe them, together with higher-index
	counterparts obtained by further differentiation,
	as $u$-tensors.
	They inherit
	a dependence on $\vect{x}$ through evaluation at
	$\phi_\alpha = \phi_\alpha(\vect{x})$.
	After transformation to Fourier space,
	the subtractions
	in~Eq.~\eqref{eq:real-deviation}
	correspond to discarding disconnected correlation functions.
	Therefore
	statistical properties of the ensemble 
	do not
	depend on our choice of fiducial point.
	
	If $\coscale/\smoothscale$ is not superexponentially large,
	we can typically expect $|\delta \phi_\alpha(\vect{r})|$ to be
	small and slowly varying.
	In Fourier space, this implies that
	$\delta \phi_\alpha(\vect{r})$
	is constructed from only a few soft, infrared modes
	which we label $\vect{k}$.
	The remaining modes have been integrated out in the smoothing process
	used to obtain this effective, separate-universe description.
	Working explicitly in terms of these modes, Eq.~\eqref{eq:real-deviation}
	yields a connecting vector and deviation equation
	for each combination of species and $\vect{k}$-mode%
		\footnote{In Eq.~\eqref{eq:k-deviation} we are keeping nonlinear
		terms in the evolution equation.
		We use the term ``Jacobi field'' to refer
		to infinitesimal connecting vectors, for which only the linear
		term need be kept.}
	\begin{equation}
		\frac{\d \delta\phi_{\kindex{\alpha}}}{\d N}
		= u_{\kindex{\alpha} \kindex{\beta}}(\vect{x})
			\delta \phi_{\kindex{\beta}}
		+ \frac{1}{2}
			u_{\kindex{\alpha} \kindex{\beta} \kindex{\gamma}}(\vect{x})
			\left\{
				\delta \phi_{\kindex{\beta}} \delta \phi_{\kindex{\gamma}}
				-
				\langle
					\delta\phi_{\kindex{\beta}}
					\delta\phi_{\kindex{\gamma}}
				\rangle
			\right\}
		+ \cdots .
		\label{eq:k-deviation}
	\end{equation}
	\end{widetext}\label{primed-index-convention}
	Eq.~\eqref{eq:k-deviation} has been written in an abbreviated
	``de Witt'' notation, in which the primed, compound index $\kindex{\alpha}$
	carries \emph{both} an unprimed species (or ``flavour'') label
	$\alpha$ and a momentum $\vect{k}_\alpha$.
	Contraction over primed indices implies summation over the flavour
	label and integration over the momentum label with measure
	$\d^3 k$.
	The 2- and 3-index $u$-tensors appearing here satisfy
	\begin{subequations}
	\begin{align}
		u_{\kindex{\alpha} \kindex{\beta}}(\vect{x}) & \equiv
			\delta(\vect{k}_\alpha - \vect{k}_\beta) u_{\alpha\beta}(\vect{x})
		\label{eq:u-two}
		\\
		u_{\kindex{\alpha} \kindex{\beta} \kindex{\gamma}}(\vect{x}) & \equiv
			(2\pi)^{-3}
			\delta(\vect{k}_\alpha - \vect{k}_\beta - \vect{k}_\gamma)
			u_{\alpha\beta\gamma}(\vect{x}) .
		\label{eq:u-three}
	\end{align}
	\end{subequations}

	Eq.~\eqref{eq:real-deviation} was given by
	Yokoyama~{\etal}~\cite{Yokoyama:2007uu, *Yokoyama:2007dw}
	in real space, and used to obtain
	evolution equations for the momentum-independent
	Lyth--Rodr\'{\i}guez Taylor coefficients.
	We explore the relationship between our approaches in
	Appendix~\ref{appendix:backwards}.
	However, Yokoyama~{\etal} did not interpret $u_{\alpha\beta}$ as the
	expansion tensor of the flow
	or give the $\vect{k}$-space
	equations~\eqref{eq:k-deviation}
	and~\eqref{eq:u-two}--\eqref{eq:u-three}.
	As we will see, this $\vect{k}$-dependent information
	is necessary to obtain transport equations for the
	full set of coupled
	$\vect{k}$-space correlation functions.
	
	One can arrive at the same conclusions using cosmological perturbation
	theory.
	Taking the background value of $\phi_\alpha$ to be the average
	field over
	the $\coscale$-sized box,
	the perturbations
	within the box
	are $\delta \phi_\alpha(\vect{r})$. One should
	now interpret
	$\vect{r}$ as a coordinate relative to the box.
	Restricting attention to the infrared modes in
	$\delta \phi_\alpha(\vect{r})$,
	for which $k/a$ is negligible,
	we recover Eqs.~\eqref{eq:k-deviation}
	and~\eqref{eq:u-two}--\eqref{eq:u-three}.
	
	\para{Correlation functions}%
	The full set of connecting vectors
	contains all information required to determine evolution of the
	bundle,
	and therefore the evolution of all statistical quantities.
	In Eq.~\eqref{eq:k-deviation} this data
	is carried by the $u$-tensors.
	The transport equations for correlation functions
	are simply a reorganization of this information.
	Therefore
	they must also be expressible
	purely in terms of $u$-tensors.
	Since~\eqref{eq:k-deviation}
	shows that these tensors
	can be obtained by the separate universe
	argument or traditional perturbation theory, it follows that they
	will make equivalent predictions.
	
	There are multiple ways to
	organize the $u$-tensors to produce evolution equations.
	In Ref.~\cite{Mulryne:2009kh,*Mulryne:2010rp},
	transport equations were obtained
	after postulating a conservation equation for a probability density $P$,
	\begin{equation}
	\label{eq:prob-conv}
		\frac{\d P}{\d N} + \partial_\alpha ( u_\alpha P ) = 0 .
	\end{equation}
	Evolution equations for the moments of $P$
	were extracted using
	both a Gauss--Hermite expansion, and generating functions.
	Here we describe a third, simpler method. Provided the perturbations
	can be treated classically,
	we expect $\d \langle O \rangle / \d N = \langle \d O / \d N
	\rangle$ for any quantity $O$.%
		\footnote{This equation both implies and is implied by
		conservation of probability, Eq.~\eqref{eq:prob-conv}}

	\subpara{Two-point function}%
	We write the two-point function
	as $\Sigma_{\kindex{\alpha} \kindex{\beta}}
	\equiv \langle \delta \phi_{\kindex{\alpha}}
	\delta \phi_{\kindex{\beta}} \rangle$.
	Eq.~\eqref{eq:k-deviation} implies
	\begin{equation}
	\begin{split}
		\frac{\d \Sigma_{\kindex{\alpha} \kindex{\beta}}}{\d N}
		& =
		\left\langle
			\frac{\d \delta\phi_{\kindex{\alpha}}}{\d N}
			\delta \phi_{\kindex{\beta}}
			+
			\delta \phi_{\kindex{\alpha}}
			\frac{\d \delta\phi_{\kindex{\beta}}}{\d N}
		\right\rangle
		\\
		& = u_{\kindex{\alpha} \kindex{\gamma}}
			\Sigma_{\kindex{\gamma} \kindex{\beta}}
			+ u_{\kindex{\beta} \kindex{\gamma}}
			\Sigma_{\kindex{\gamma} \kindex{\alpha}}
		+ \text{[$\geq$ 3 p.f.]}
		\\
		& \equiv \{ u , \Sigma \}_{\kindex{\alpha}\kindex{\beta}}
		+ \text{[$\geq$ 3 p.f.]}
	\end{split}
	\label{eq:2pf-transport}
	\end{equation}
	where $\{ A, B \}$ is the matrix anticommutator of $A$ and $B$,
	and
	[$\geq$ 3 p.f.] denotes
	terms including three-point functions or above which have
	been omitted.
	In general, the transport equations will couple correlation functions
	of all orders.
	They can be thought of as a limiting case
	of a Schwinger--Dyson hierarchy,
	applied to
	expectation values 
	rather than
	the \emph{in--out} amplitudes of scattering theory.
	Calzetta \& Hu argued that the result could be interpreted as
	a Boltzmann hierarchy \cite{Calzetta:1996sy,Calzetta:1999xh}.
	
	As in any effective theory,
	the transport equations will be useful only if
	a reason can be found to systematically neglect
	an infinite number of terms.
	Applied to inflation,
	the statistical properties of the ensemble are nearly Gaussian:
	in the simplest models, an $n$-point function
	will typically be of order $H^{m(n)}$, where $m(n)$
	is the smallest even integer at least as large as $n$~\cite{Jarnhus:2007ia}.
	This is suppressed compared to the natural scale $\Mp$ by
	$(H/\Mp)^{m(n)} \ll 1$.
	However, this is not necessary;
	all that is required (or suggested by observation)
	for~\eqref{eq:2pf-transport}
	to be valid is that the
	three- and higher $n$-point functions are substantially smaller than
	the two-point function.
	
	Eq.~\eqref{eq:2pf-transport}
	was given in Ref.~\cite{Mulryne:2009kh,*Mulryne:2010rp}
	for an arbitrary $n$-field model,
	but with $\Sigma_{\kindex{\alpha} \kindex{\beta}}$
	interpreted as the real-space correlation function.
	The single-field case is discussed by Gardiner~\cite{citeulike:1400625}.
	With the $u$-tensors given in~\eqref{eq:u-two}--\eqref{eq:u-three},
	Eq.~\eqref{eq:2pf-transport}
	applies for the full $\vect{k}$-dependent
	correlation function.
	
	\subpara{Three-point function}%
	We write
	the three-point function
	as $\alpha_{\kindex{\alpha} \kindex{\beta} \kindex{\gamma}}
	\equiv \langle
	\delta \phi_{\kindex{\alpha}}
	\delta \phi_{\kindex{\beta}}
	\delta \phi_{\kindex{\gamma}} \rangle$.
	Keeping contributions of order $\Or(\Sigma^2)$ and $\Or(\alpha)$,
	we conclude
	\begin{equation}
	\begin{split}
		\frac{\d \alpha_{\kindex{\alpha} \kindex{\beta} \kindex{\gamma}}}{\d N}
		= \mbox{} &
		u_{\kindex{\alpha} \kindex{\lambda}}
		\alpha_{\kindex{\lambda} \kindex{\beta} \kindex{\gamma}}
		+
		u_{\kindex{\alpha} \kindex{\lambda} \kindex{\mu}}
		\Sigma_{\kindex{\lambda} \kindex{\beta}}
		\Sigma_{\kindex{\mu} \kindex{\gamma}}
		\\ & \mbox{}
		+
		\text{cyclic ($\kindex{\alpha} \rightarrow \kindex{\beta}
		\rightarrow \kindex{\gamma}$)}
		+
		\text{[$\geq$ 4 p.f.]} .
	\end{split}
	\label{eq:3pf-transport}
	\end{equation}
	In simple models,
	the scaling estimate $\langle \delta \phi^n \rangle \sim H^{m(n)}$
	makes both terms the same order of magnitude.
	For~\eqref{eq:3pf-transport} to be valid requires
	the 4-point function to be substantially smaller,
	which is also supported by observation
	\cite{Smidt:2010ra,*Fergusson:2010gn}.
	
	\section{Evolution of correlation functions}
	\label{sec:evolution}

	\subsection{Solution of the transport hierarchy by raytracing}
	\label{sec:raytracing-solution}
	
	The transport equations~\eqref{eq:2pf-transport}
	and~\eqref{eq:3pf-transport} can be solved
	using the machinery developed in~\S\ref{sec:geo-optics}.
	The key ingredients are
	the phase-space flows which describe evolution of individual ``separate
	universes,'' and the Jacobi fields which connect them.
	The solution is formal and depends only on the structure
	described in \S\ref{sec:transport}.
	Therefore there is no requirement to impose the slow-roll
	approximation,
	and when
	written over the full phase-space our equations
	apply quite generally.
	When truncated to field-space they reproduce the
	slow-roll evolution.
		
	\para{Two-point function}%
	We write the two-point function $\Sigma_{\kindex{\alpha} \kindex{\beta}}$
	in the form
	\begin{equation}
		\Sigma_{\kindex{\alpha}\kindex{\beta}} \equiv
			\Gamma_{\kindex{\alpha} \kindex{i}}
			\Gamma_{\kindex{\beta} \kindex{j}}
			\Sigma_{\kindex{i} \kindex{j}} ,
		\label{eq:2pf-solution}
	\end{equation}
	where $\Gamma$ is to be determined.
	This notation has been chosen because $\Gamma$ will turn out to be the
	propagator matrix~\eqref{eq:propagator} for the primed indices.
	Indeed,~\eqref{eq:2pf-solution}
	is a solution of the transport equation~\eqref{eq:2pf-transport}
	if
	\begin{subequations}
	\begin{align}
		\frac{\d \Gamma_{\kindex{\alpha}\kindex{i}}}{\d N}
		& =
		u_{\kindex{\alpha} \kindex{\gamma}}
		\Gamma_{\kindex{\gamma}\kindex{i}}
		\label{eq:k-propagator} \\
		\frac{\d \Sigma_{\kindex{i} \kindex{j}}}{\d N}
		& = 
		\Or( H^4 )
		\approx 0 .
		\label{eq:2pf-kernel}
	\end{align}
	\end{subequations}
	Eq.~\eqref{eq:k-propagator}
	is the equation for a Jacobi field, Eq.~\eqref{eq:deviation-equation}.
	
	In writing~\eqref{eq:2pf-kernel} we have assumed
	approximate Gaussianity, so
	that contributions from higher-order correlation
	functions are suppressed by at least a power of $H^2$
	compared to the terms which have been retained.
	Keeping these terms would yield the ``loop corrections''
	of the Lyth--Rodr\'{\i}guez formalism
	\cite{Boubekeur:2005fj,Lyth:2006qz,Seery:2007we,*Seery:2007wf,Seery:2010kh}.
	To the order we are working,
	$\Sigma_{\kindex{i}\kindex{j}}$ should be identified as a constant:
	it is the value of the two-point function
	evaluated at $N=N_0$,
	where $N_0$ is the initial time which appears in the
	propagator~\eqref{eq:propagator}.
	We write this constant value
	$\Sinit_{\kindex{i}\kindex{j}}$.
	
	The primed propagator satisfies
	\begin{equation}
		\Gamma_{\kindex{\alpha} \kindex{i}}
		=
		\delta(\vect{k}_\alpha - \vect{k}_i)
		\Gamma_{\alpha i} ,
		\label{eq:k-propagator-solution}
	\end{equation}
	where $\Gamma_{\alpha i}$ is the flavour propagator~\eqref{eq:propagator}.
	Therefore, written more explicitly, Eq.~\eqref{eq:2pf-solution}
	becomes
	\begin{equation}
		\langle
		\delta \phi_{\alpha}(\vect{k}_\alpha)
		\delta \phi_{\beta}(\vect{k}_\beta)
		\rangle
		=
		\Gamma_{\alpha i}
		\Gamma_{\beta j}
		\langle
		\delta \phi_i(\vect{k}_\alpha)
		\delta \phi_j(\vect{k}_\beta)
		\rangle_0 ,
		\label{eq:2pf-explicit}
	\end{equation}
	where our usual convention---that Latin indices denote evaluation
	of the correlation function at some initial time $N_0$---%
	continues to apply.
	For the two point function, practical calculations
	usually simplify if this is taken to be the horizon-crossing time
	associated with scale $k = |\vect{k}_\alpha| = |\vect{k}_\beta|$.
	We have indicated this by attaching a subscript `$0$'
	to the correlation function.
	With this
	understanding,
	and recollecting the identification~\eqref{eq:gamma-deriv},
	Eq.~\eqref{eq:2pf-explicit} is
	the familiar ``$\delta N$'' result
	\cite{Starobinsky:1986fxa,Sasaki:1995aw,Lyth:2005fi}.
	
	\para{Three-point function}%
	Similar methods can be used to solve for the three- and four-point
	functions.
	We write
	$\alpha_{\kindex{\alpha}\kindex{\beta}\kindex{\gamma}} \equiv
	\Gamma_{\kindex{\alpha}\kindex{i}}
	\Gamma_{\kindex{\beta}\kindex{j}}
	\Gamma_{\kindex{\gamma}\kindex{k}}
	\alpha_{\kindex{i}\kindex{j}\kindex{k}}$.
	As for the two-point function, the propagator matrices
	absorb contributions from the $u_{\alpha\beta}$-tensors.
	In the case of $\Sigma_{\kindex{\alpha}\kindex{\beta}}$ there were
	no other terms, making the ``kernel'' $\Sigma_{\kindex{i}\kindex{j}}$
	time independent.
	Here, the presence of terms involving $u$ 3-tensors
	provides a source for
	$\alpha_{\kindex{i}\kindex{j}\kindex{k}}$.
	We find%
		\footnote{We are allowing $\alpha_{\kindex{i}\kindex{j}\kindex{k}}$
		to be a function of $N$,
		which means our index convention must be interpreted
		more abstractly.
		The expressions for $\Gamma$-matrices to which
		Eq.~\eqref{eq:3pf-kernel} leads, such as
		Eqs.~\eqref{eq:dn1}--\eqref{eq:dn2},
		can be interpreted in the original sense.}
	\begin{equation}
		\begin{split}
		\frac{\d \alpha_{\kindex{i}\kindex{j}\kindex{k}}}{\d N} = \mbox{}
		&
		(\Gamma^{-1}_{\kindex{i}\kindex{\alpha}}
		u_{\kindex{\alpha}\kindex{\beta}\kindex{\gamma}}
		\Gamma_{\kindex{\beta}\kindex{m}}
		\Gamma_{\kindex{\gamma}\kindex{n}})
		\Sinit_{\kindex{m}\kindex{j}}
		\Sinit_{\kindex{n}\kindex{k}}
		+ \text{cyclic}
		\\ & \mbox{}
		+ \Or(H^6) ,
		\end{split}
		\label{eq:3pf-kernel}
	\end{equation}
	where, as above,
	$\Sinit_{ij}$ is the initial value of the two-point function introduced
	in~\eqref{eq:2pf-solution}.
	The estimate $\Or(H^6)$ for the truncation error, beginning with
	contributions from the four-point function, again assumes
	that the correlation functions order themselves in even powers of $H$.
	We define the matrix $\Gamma^{-1}_{\kindex{i} \kindex{\alpha}}$
	to be the left-inverse of the propagator,
	$\Gamma^{-1}_{\kindex{i} \kindex{\alpha}}
	\Gamma_{\kindex{\alpha} \kindex{j}}
	= \delta(\vect{k}_i - \vect{k}_j) \delta_{ij}$.
	Inspection of~\eqref{eq:k-propagator-solution} shows that it
	can be written
	\begin{equation}
		\Gamma^{-1}_{\kindex{i} \kindex{\alpha}}
		= \delta(\vect{k}_i - \vect{k}_\alpha) \Gamma^{-1}_{i \alpha} ,
		\label{eq:k-inverse-propagator-solution}
	\end{equation}
	where $\Gamma^{-1}_{i \alpha}$ is the conventional matrix inverse
	of the flavour propagator, Eq.~\eqref{eq:propagator}.
	In what follows it is useful to define a projected $u$ 3-tensor,
	$\tilde{u}_{\kindex{i} \kindex{j} \kindex{k}}$, by
	\begin{equation}
		\tilde{u}_{\kindex{i}\kindex{j}\kindex{k}}
			=
			\Gamma^{-1}_{\kindex{i}\kindex{\alpha}}
			u_{\kindex{\alpha}\kindex{\beta}\kindex{\gamma}}
			\Gamma_{\kindex{\beta}\kindex{j}}
			\Gamma_{\kindex{\gamma}\kindex{k}} .
	\end{equation}
	Combining~\eqref{eq:k-propagator-solution}
	and~\eqref{eq:k-inverse-propagator-solution},
	it follows that the explicit $\vect{k}$- and flavour-dependence can be
	written
	\begin{equation}
		\tilde{u}_{\kindex{i}\kindex{j}\kindex{k}}
		=
		\delta(\vect{k}_i - \vect{k}_j - \vect{k}_k)
		\tilde{u}_{ijk} ,
	\end{equation}
	where the tensor $\tilde{u}_{ijk}$ is the obvious flavour projection
	of $u_{ijk}$, so that
	$\tilde{u}_{ijk} = \Gamma^{-1}_{i\alpha} u_{\alpha\beta\gamma}
	\Gamma_{\beta j}\Gamma_{\gamma k}$.
	
	With these definitions, Eq.~\eqref{eq:3pf-kernel} can be solved by
	quadrature. Up to loop corrections, we find
	\begin{equation}
		\alpha_{\kindex{i}\kindex{j}\kindex{k}}
		=
			\Ainit_{\kindex{i}\kindex{j}\kindex{k}}
			+ \int_{N_0}^N \tilde{u}_{\kindex{i}\kindex{m}\kindex{n}}(\varN)
			\Sinit_{\kindex{m}\kindex{j}}
			\Sinit_{\kindex{n}\kindex{k}} \; \d \varN
			+ \text{cyclic} ,
		\label{eq:3pf-solution}
	\end{equation}
	where
	$\Ainit_{\kindex{i}\kindex{j}\kindex{k}}$
	should be regarded as the value of the three-point
	function at $N = N_0$.
	The complete solution can be written
	(again up to loop corrections)
	\begin{equation}
	\begin{split}
		\alpha_{\kindex{\alpha}\kindex{\beta}\kindex{\gamma}} = \mbox{} &
			\Gamma_{\kindex{\alpha}\kindex{i}}
			\Gamma_{\kindex{\beta}\kindex{j}}
			\Gamma_{\kindex{\gamma}\kindex{k}}
			\Ainit_{\kindex{i}\kindex{j}\kindex{k}}
			\\ &
			\mbox{}
			+
			\Big(
				\Gamma_{\kindex{\alpha}\kindex{m}\kindex{n}}
				\Gamma_{\kindex{\beta}\kindex{j}}
				\Gamma_{\kindex{\gamma}\kindex{k}}
				\Sinit_{\kindex{m}\kindex{j}}
				\Sinit_{\kindex{n}\kindex{k}}
				+
				\text{cyclic}
			\Big) ,
	\end{split}
	\label{eq:3pf-total}
	\end{equation}
	where the cyclic permutations
	exchange $\kindex{\alpha} \rightarrow \kindex{\beta} \rightarrow
	\kindex{\gamma}$.

	One can regard Eqs.~\eqref{eq:k-propagator}--\eqref{eq:2pf-kernel}
	and~\eqref{eq:3pf-solution}
	as analogous to the ``line of sight'' integral which is used to
	obtain a formal solution to the Boltzmann equation
	in calculations of the cosmic microwave background anisotropies.

	The quantity $\Gamma_{\kindex{\alpha}\kindex{m}\kindex{n}}$
	is defined by
	\begin{equation}
		\Gamma_{\kindex{\alpha}\kindex{m}\kindex{n}} \equiv
			\Gamma_{\kindex{\alpha}\kindex{i}}
			\int_{N_0}^N
			\tilde{u}_{\kindex{i}\kindex{m}\kindex{n}}(\varN) \; \d \varN .
		\label{eq:gamma-2}
	\end{equation}
	Observe that
	Eq.~\eqref{eq:gamma-2} is symmetric in the indices
	$\kindex{m}$ and $\kindex{n}$.
	With our choices for the $\vect{k}$- and flavour-dependence
	of its constituent quantities,
	it can be written
	\begin{equation}
		\Gamma_{\kindex{\alpha}\kindex{m}\kindex{n}} =
		\delta(\vect{k}_\alpha - \vect{k}_m - \vect{k}_n)
		\Gamma_{\alpha m n} ,
	\end{equation}
	where $\Gamma_{\alpha m n }$ is the flavour-only object obtained by
	exchanging primed for unprimed indices in~\eqref{eq:gamma-2}.
	Comparing with~\eqref{eq:propagator}, it follows that
	(up to matrix ordering ambiguities)
	$\Gamma_{\alpha m n}$ is the derivative of the propagator,
	\begin{equation}
		\label{eq:second-deriv}
		\frac{\partial^2 \phi_\alpha}{\partial \phi_m \partial \phi_n}
		= \Gamma_{\alpha m n} .
	\end{equation}
	Eq.~\eqref{eq:3pf-total} can now be recognized as the
	Lyth--Rodr\'{\i}guez formula for the three-point function
	\cite{Lyth:2005fi}.

	\subsection{Flow equations for ``$\delta N$'' coefficients}
	\label{sec:flow-deltaN}
	
	We conclude that the transport equations~\eqref{eq:2pf-transport}
	and~\eqref{eq:3pf-transport} are equivalent to
	the Taylor expansion algorithm of Lyth \& Rodr\'{\i}guez
	for the three-point function.
	Also, because the $u$-tensors could equally well be derived
	using the methods of cosmological perturbation theory,
	all these methods will give answers which agree.
	Within this narrow reading,
	our analysis can be interpreted as a demonstration
	that these methods are interchangeable.
	Therefore
	we believe that statements
	to the effect that any particular
	method currently in use has an intrinsic drawback
	when compared with another,
	\emph{as a matter of principle},
	are wrong.
	
	Nevertheless it \emph{is} true that some approaches
	have advantages in practice,
	although no one approach outperforms the others
	in all applications.
	For example,
	as explained in \S\ref{sec:introduction},
	in some models the Taylor expansion
	algorithm leads to very simple analytic formulae.
	This property has encouraged a large literature
	studying
	models to which the method can be applied.

	In this broader context 
	our analysis is not simply a
	reformulation of existing results.
	First,
	as a byproduct of the raytracing method we have obtained explicit
	(but formal) expressions for
	the Lyth--Rodr\'{\i}guez Taylor coefficients,
	\begin{subequations}
	\begin{align}
		\frac{\partial \phi_\alpha}{\partial \phi_i}
			& = \Gamma_{\alpha i}
				=
				\pathorder \exp
				\left(
					\int_{N_0}^{N} u_{\alpha\beta}(\varN) \; \d \varN
				\right)
				\delta_{\beta i}
			\label{eq:dn1}
			\\
		\frac{\partial^2 \phi_\alpha}{\partial \phi_i \partial \phi_j}
		 	& = \Gamma_{\alpha ij}
		 		=
		 		\Gamma_{\alpha m}
		 		\int_{N_0}^N \tilde{u}_{mij}(\varN) \; \d \varN .
		 	\label{eq:dn2}
	\end{align}
	\end{subequations}
	Analytically, the Taylor expansion method is useful only
	when a solution to~\eqref{eq:dn1} can be found in closed form.
	This has been achieved only for a limited class of potentials
	obeying some form of separability criteria;
	a summary appears in Ref.~\cite{Elliston:2011dr}
	together with references to the original literature.
	Eq.~\eqref{eq:dn1} clarifies the difficulty encountered in obtaining
	analytic formulae as the
	difficulty of computing closed-form
	expressions for a path-ordered exponential.
	A sophisticated theory is available
	\cite{Lam:1998tw}
	but explicit expressions can usually be obtained only in special
	cases, or where the expansion tensor commutes with itself
	at different times.
	It is possible that Eq.~\eqref{eq:dn1} could be used
	to extend analytic progress
	beyond the separable cases, but we have not investigated this
	possibility in detail. 

	Eqs.~\eqref{eq:dn1}--\eqref{eq:dn2} were given,
	in slightly different notation, by
	Yokoyama~{\etal}~\cite{Yokoyama:2007uu, *Yokoyama:2007dw}.
	Because of its close relation 
	to the present discussion
	we review and extend the
	Yokoyama~{\etal} approach in Appendix~\ref{appendix:backwards}.
	
	Second, a na\"{\i}ve numerical implementation
	of the Taylor expansion
	formula is unfavourable.
	Beginning with fractionally displaced initial conditions one must evolve
	the equations of motion over many e-folds, during
	which numerical noise is accumulating.
	Taking differences between these evolved solutions
	requires high-accuracy integration in order that the small displacement
	in initial conditions is not swamped by noise.
	The explicit solutions~\eqref{eq:dn1}--\eqref{eq:dn2}
	allow this na\"{\i}ve approach to be replaced
	by a simple system of ordinary differential equations for
	$\Gamma_{\alpha i}$ and $\Gamma_{\alpha ij}$.
	The $\Gamma_{\alpha i}$ equation is the Jacobi
	equation~\eqref{eq:k-propagator},
	after dropping primes on indices.
	The initial condition is $\Gamma_{\alpha i} = \delta_{\alpha i}$.
	The $\Gamma_{\alpha ij}$ equation can be obtained
	by differentiation of~\eqref{eq:dn2}. It is
	\begin{equation}
		\label{eq:gamma2-evolve}
		\frac{\d \Gamma_{\alpha ij}}{\d N}
		=
			u_{\alpha \beta} \Gamma_{\beta i j}
			+ u_{\alpha \beta \gamma} \Gamma_{\beta i} \Gamma_{\gamma j} ,
	\end{equation}
	with initial condition $\Gamma_{\alpha ij} = 0$.
	
	The same approach can be applied systematically to deduce transport
	equations for any of the Taylor coefficients.
	Yokoyama~{\etal} wrote the transport equation~\eqref{eq:k-propagator}
	for $\Gamma_{\alpha i}$, but did not write~\eqref{eq:gamma2-evolve}
	for $\Gamma_{\alpha ij}$
	which they computed directly from~\eqref{eq:dn2}.
	See Appendix~\ref{appendix:backwards}
	for a comparison.
	
	\subsection{Transport of ``shape'' amplitudes}
	\label{sec:shape}
	
	The results of \S\ref{sec:raytracing-solution} apply for arbitrary
	initial conditions
	$\Sinit_{\kindex{i}\kindex{j}}$,
	$\Ainit_{\kindex{i}\kindex{j}\kindex{k}}$
	for the two- and three-point functions.
	But for application to inflation, we will usually wish to
	apply them to the correlation functions produced in a specific
	model. In this case the fields $\phi_\alpha$
	will be a collection of light scalars
	for which
	$\Sinit_{\kindex{i}\kindex{j}}$ and
	$\Ainit_{\kindex{i}\kindex{j}\kindex{k}}$
	can be computed using the \emph{in--in} formulation of quantum
	field theory
	\cite{Seery:2005gb}.
	These yield very specific $\vect{k}$-dependences whose amplitudes
	we wish
	to track.
	
	In this section, our analysis remains general and
	continues to apply
	to the full phase space.	
	
	\para{Two-point function}%
	The two-point function is straightforward.
	For a nearly scale-invariant spectrum we
	have
	\begin{equation}
		\Sigma_{\kindex{\alpha}\kindex{\beta}}
		\equiv
		(2\pi)^3 \delta(\vect{k}_\alpha + \vect{k}_\beta)
		\frac{\Sigma_{\alpha\beta}}{k^3} ,
		\label{eq:2pf-k}
	\end{equation}
	where $k = |\vect{k}_\alpha| = |\vect{k}_\beta|$
	and the flavour matrix $\Sigma_{\alpha\beta}$
	should be nearly independent of $k$.
	Transport of $\Sigma_{\alpha\beta}$ can be
	accomplished using~\eqref{eq:2pf-explicit}, or simply by solving
	the transport equation~\eqref{eq:2pf-transport}
	with an appropriate initial condition
	after dropping primes on indices.
	That gives
	\begin{equation}
		\Sigma_{\alpha\beta} = \Gamma_{\alpha i} \Gamma_{\beta j}
		\Sinit_{ij} ,
		\label{eq:sigma-flavour}
	\end{equation}
	where $\Sinit_{ij}$ is the initial value of $\Sigma_{\alpha\beta}$.
	The mild $k$-dependence of~\eqref{eq:sigma-flavour}
	can also be obtained using transport techniques~\cite{Dias:2011xy}.
	
	\para{Three-point function}%
	Here, more possibilities exist.
	It is known that the
	$\Or(\Sinit^2)$ terms in~\eqref{eq:3pf-total}
	dominate whenever the bispectrum is large enough to be
	observed~\cite{Lyth:2005qj,Vernizzi:2006ve}.
	Eq.~\eqref{eq:3pf-total} shows that these contributions add
	incoherently to the contribution from $\Ainit_{ijk}$, so they can be
	studied separately.
	Using~\eqref{eq:2pf-k} and overall symmetry of the correlation function
	under exchange of indices, we can write
	\begin{equation}
		\alpha_{\kindex{\alpha}\kindex{\beta}\kindex{\gamma}}
		\supseteq
		(2\pi)^3 \delta(\vect{k}_\alpha + \vect{k}_\beta + \vect{k}_\gamma)
		\left(
			\frac{\alpha_{\alpha\mid\beta\gamma}}{k_\beta^3 k_\gamma^3}
			+ \frac{\alpha_{\beta\mid\alpha\gamma}}{k_\alpha^3 k_\gamma^3}
			+ \frac{\alpha_{\gamma\mid\alpha\beta}}{k_\alpha^3 k_\beta^3}
		\right) ,
		\label{eq:3pf-shape}
	\end{equation}
	where the notation ``$\supseteq$'' indicates that the three-point
	contribution contains this contribution among others.
	The amplitudes $a_{\alpha\mid\beta\gamma}$ are symmetric under
	exchange of $\beta$ and $\gamma$, but not otherwise.
	Using Eqs.~\eqref{eq:u-two}, \eqref{eq:u-three},
	\eqref{eq:3pf-transport}
	and~\eqref{eq:2pf-k}, we find
	the transport equation
	\begin{equation}
	\begin{split}
		\frac{\d \alpha_{\alpha\mid\beta\gamma}}{\d N} = \mbox{} &
			u_{\alpha \lambda} \alpha_{\lambda\mid\beta\gamma}
			+ u_{\beta \lambda} \alpha_{\alpha\mid\lambda\gamma}
			+ u_{\gamma \lambda} \alpha_{\alpha\mid\beta\lambda}
		\\ & \mbox{}
			+ u_{\alpha \lambda\mu} \Sigma_{\lambda\beta} \Sigma_{\mu\gamma} .
	\end{split}
	\label{eq:alpha-flavour-transport}
	\end{equation}
	If desired,
	we can apply the same method of formal solution
	described in \S\ref{sec:raytracing-solution}.
	This yields
	\begin{equation}
		\alpha_{\alpha\mid\beta\gamma} =
			\Gamma_{\alpha mn} \Gamma_{\beta j} \Gamma_{\gamma k}
			\Sinit_{mj} \Sinit_{nk} .
		\label{eq:a-solution}
	\end{equation}
	In combination with~\eqref{eq:3pf-shape} this
	reproduces
	our earlier formula~\eqref{eq:3pf-total}, neglecting the initial
	contribution $\Ainit_{\kindex{i}\kindex{j}\kindex{k}}$.
	
	\subsection{Connections between the transport and other approaches}
	\label{sec:connections}
	
	Up to this point we have
	shown
	that the Jacobi fields which connect ``separate universe''
	trajectories in phase space can be used to solve the transport
	equations for the full set of $\vect{k}$-space
	correlation functions.
	But as we have explained, the transport hierarchy is just one of
	many techniques for handling
	correlation functions.
	We now pause to examine the connections
	between these approaches.

	\para{$\delta N$ formalism}%
	In the Lyth--Rodr\'{\i}guez approach, or
	``$\delta N$ formalism'',
	one makes a Taylor expansion of the field values on a final
	hypersurface in terms of field values on some initial hypersurface.
	Following the discussion surrounding Eq.~\eqref{eq:real-deviation},
	and with the same meaning for the vectors $\vect{x}$ and $\vect{r}$,
	this can be written
	\begin{widetext}
	\begin{equation} 
	\label{eq:taylor}
		\delta \phi_\alpha(\vect{r}) =
		\Gamma_{\alpha i}(\vect{x}) \delta \phi_i(\vect{r}) + 
		\frac{1}{2} \Gamma_{\alpha i j}(\vect{x})
			\left\{
				\delta \phi_i(\vect{r}) \delta \phi_j(\vect{r})
				-	
				\langle
					\delta \phi_i(\vect{r}) \delta \phi_j(\vect{r})
				\rangle
			\right\}
		+ \dots .
	\end{equation}
	\end{widetext}
	Note that, despite appearances, we are making no assumption that
	the evolution of $\delta \phi$ is close to an attractor.
	Therefore there is no requirement to invoke the slow-roll approximation.
	It is true that the existence of an attractor
	would make the canonical momenta purely a function of the
	fields, yielding an equation with the
	appearance of Eq.~\eqref{eq:taylor}.
	But
	as we have explained, by working in a first-order
	Hamiltonian formalism we can obtain expressions such as~\eqref{eq:taylor}
	without this limitation.
	Therefore we allow the $\delta \phi_i$ to include perturbations
	of the canonical momenta if necessary, in which case the indices
	$\alpha$, $i$, etc. range over the $2M$ dimensions of phase space.
	Where slow-roll is a good approximation we can revert to a simpler
	formulation based on field space.
	
	We have already remarked that
	the $\Gamma$-tensors are the derivatives~\eqref{eq:gamma-deriv}
	and \eqref{eq:second-deriv}.
	In Eq.~\eqref{eq:taylor} the $\delta \phi_\alpha$ are all defined on
	spatially flat hypersurfaces.
	More commonly, an analogous expansion is made for the
	total e-folding number $N$, measured from a
	flat slice to a final comoving slice;
	we give an explicit relation in~\S\ref{sec:gauge}.
	The choice of slicing simply corresponds to the gauge in which 
	we wish to work \cite{Mulryne:2010rp}.

	For~\eqref{eq:taylor} to be useful, some means must be found
	to compute $\Gamma_{\alpha i}$ and $\Gamma_{\alpha i j}$.

	\para{Flow equations}%
	As a by-product of the raytracing solution, or
	``line of sight'' integral,
	we obtained the evolution equations~\eqref{eq:k-propagator}
	and \eqref{eq:gamma2-evolve}.
	These allow the $\Gamma$-tensors to be computed easily.
	However, the same equations can be obtained directly from
	the separate universe formula,
	Eq.~\eqref{eq:taylor}.
	Substituting~\eqref{eq:taylor} into 
	both the right- and left-hand sides of~\eqref{eq:real-deviation}
	and separating 
	the resulting expansion order-by-order, we immediately 
	arrive at Eqs.~\eqref{eq:k-propagator} and \eqref{eq:gamma2-evolve}.
	This still does not require the slow-roll approximation.

	\para{Transfer matrices}%
	We have observed that
	Eq.~\eqref{eq:real-deviation}
	arises in the $k/aH \rightarrow 0$ limit of
	cosmological perturbation theory (``CPT'').
	Within that 
	framework, at least in the first-order theory,
	it is common to introduce ``transfer matrices''
	which relate
	field perturbations at different times~\cite{Amendola:2001ni}.
	Typically these are chosen to be
	the adiabatic and isocurvature directions,
	but in principle any basis can be used.

	Restricting to first-order,
	the transfer matrix
	is determined precisely by the leading term of~\eqref{eq:taylor},
	or a gauge transformation of it.
	It follows that
	Eq.~\eqref{eq:taylor} represents the extension
	of the transfer matrix to second-order (and beyond),
	and Eqs.~\eqref{eq:k-propagator} and~\eqref{eq:gamma2-evolve} 
	give the evolution of the transfer tensors
	$\Gamma_{\alpha i}$, $\Gamma_{\alpha i j}$.
	Therefore the transfer-matrix formalism is precisely equivalent to
	the separate universe picture and traditional cosmological perturbation
	theory.
	Note that if the perturbations 
	are projected onto adiabatic and isocurvature modes
	this requires use of the 
	correct $u$ tensors at each time step. 
	
	\para{CPT implies transport equations}%
	Finally, we show that cosmological perturbation theory implies
	the transport hierarchy with which we began.
	We write 
	\begin{equation}
	\Sigma_{\alpha \beta} = \Gamma_{\alpha i} \Gamma_{\beta j} \Sinit_{i j}  , 	
	\end{equation}
	which, neglecting ``loops,'' follows from~\eqref{eq:taylor} and therefore
	either
	CPT or a transfer-matrix approach.
	Differentiating both sides with respect to time,
	recalling that $\Sinit_{i j}$ is time-independent,
	and make use of~\eqref{eq:k-propagator} we find 
	\begin{equation}
	\frac{\d \Sigma_{\alpha \beta}}{\d N}
		= \left (
			u_{\alpha \mu} \Gamma_{\mu i} \Gamma_{\beta j} 
			+ u_{\beta \mu} \Gamma_{\alpha i} \Gamma_{\mu j}
		\right) \Sinit_{i j} .
	\end{equation}
	This gives the transport equation
	for $\Sigma_{\alpha\beta}$, Eq.~\eqref{eq:2pf-transport}.
	A similar procedure leads to the
	transport equation for $\alpha_{\alpha\beta\gamma}$,
	Eq.~\eqref{eq:2pf-transport}.
	It follows that each of these approaches implies and is implied by
	the others.

	\section{Gauge transformations}
	\label{sec:gauge}
	
	To this point, the formalism we have developed
	enables the correlation functions of fluctuations
	in the fields and their momenta,
	$\delta \phi_\alpha$ and $\delta p_\alpha$,
	to be evolved along the bundle of trajectories
	picked out by an ensemble of smoothed regions.
	However, by themselves these fluctuations are not observable.
	Only specific combinations are observable, of which the
	most important is the primordial curvature fluctuation $\zeta$.
	Therefore to proceed we
	require expressions for the gauge transformation
	between the $\delta \phi_\alpha$, $\delta p_\alpha$
	and
	$\zeta$.
	
	In this section we impose the slow-roll
	approximation throughout, enabling us to work on field space
	and make use of the hypersurface-orthogonal property of the flow.
	We intend to return to the general case in a future
	publication.

	\subsection{Explicit transformations}
	
	In the slow-roll approximation
	there is no need to track the momentum fluctuations
	$\delta p_\alpha$, which are purely determined by
	the field fluctuations $\delta \phi_\alpha$.
	Therefore $\zeta$ can be written purely in terms of the field
	fluctuations.
	
	On superhorizon scales,
	the appropriate gauge transformation
	can be written as a Taylor expansion,
	\begin{equation}
		\zeta = N_\alpha \delta \phi_\alpha
			+ \frac{1}{2} N_{\alpha\beta} (
				\delta \phi_\alpha \delta \phi_\beta
				- \langle \delta \phi_\alpha \delta \phi_\beta \rangle
			) + \cdots ,
		\label{eq:gauge-transformation}
	\end{equation}
	where all fields are evaluated at the same spatial position
	and a constant has been subtracted to set $\langle \zeta \rangle = 0$.
	The Taylor coefficients $N_\alpha$
	and $N_{\alpha\beta}$ have been given by
	various authors	\cite{Malik:2008im,Mulryne:2009kh}.
	Working in field space,
	we give a purely geometrical derivation.
	This argument relies on the property that the flow is orthogonal
	to surfaces of constant density in field space,
	and therefore will not generalize directly to the full phase space.

	\begin{figure*}
	
		\small

		\vspace{7mm}
		\hfill
		\subfloat[][First order\label{fig:gauge-first}]{
		\begin{fmfgraph*}(150,90)
			\fmfpen{thin}
			\fmfipair{sigmaa,sigmab,sigmac}
			\fmfipair{traja,trajb,trajc}
			\fmfiequ{sigmaa}{(0,h)}
			\fmfiequ{sigmab}{(0.25w,0.5h)}
			\fmfiequ{sigmac}{(w,0)}
			\fmfiequ{traja}{point 0.75*length(sigmaa .. sigmab .. sigmac) of (sigmaa .. sigmab .. sigmac)}
			\fmfiequ{trajc}{point 0.4*length(sigmaa .. sigmab .. sigmac) of (sigmaa .. sigmab .. sigmac)}
			\fmfiequ{trajb}{(0.65w,0.8h)}
			\fmfi{plain}{sigmaa .. sigmab .. sigmac}
			\fmfi{dashes,label=$\delta \phi^1$,foreground=red}{traja .. trajb}
			\fmfi{dashes,label=$\delta \phi^{\mathrm{flow}}$,label.side=right,foreground=blue}{trajb .. trajc}
			\fmfiv{label=$\Sigma_{\rho}$,label.angle=0}{sigmac}
			\fmfiv{decor.shape=circle,decor.filled=full,decor.size=0.025w,label=$x$,label.angle=-135}{traja}
			\fmfiv{decor.shape=circle,decor.filled=full,decor.size=0.025w,label=$z$}{trajb}
			\fmfiv{decor.shape=circle,decor.filled=full,decor.size=0.025w,label=$y$,label.angle=-135}{trajc}
		\end{fmfgraph*}}
		\hfill
		\subfloat[][Second order\label{fig:gauge-second}]{
		\begin{fmfgraph*}(150,90)
			\fmfpen{thin}
			\fmfipair{sigmaa,sigmab,sigmac}
			\fmfipair{taua,taub,tauc}
			\fmfipair{traja,trajb,trajc}
			\fmfipair{flowa,flowb}
			\fmfiequ{sigmaa}{(0,h)}
			\fmfiequ{sigmab}{(0.25w,0.5h)}
			\fmfiequ{sigmac}{(w,0)}
			\fmfiequ{taua}{(0.4w,h)}
			\fmfiequ{taub}{(0.75w,0.55h)}
			\fmfiequ{tauc}{(w,0.45h)}
			\fmfiequ{flowa}{point 0.25*length(sigmaa .. sigmab .. sigmac) of (sigmaa .. sigmab .. sigmac)}
			\fmfiequ{traja}{point 0.75*length(sigmaa .. sigmab .. sigmac) of (sigmaa .. sigmab .. sigmac)}
			\fmfiequ{trajb}{point 0.6*length(taua .. taub .. tauc) of (taua .. taub .. tauc)}
			\fmfiequ{flowb}{point 0.25*length(taua .. taub .. tauc) of (taua .. taub .. tauc)}
			\fmfiequ{trajc}{(0.9w,0.9h)}
			\fmfi{plain}{sigmaa .. sigmab .. sigmac}
			\fmfi{plain}{taua .. taub .. tauc}
			\fmfi{dashes,label=$\delta \phi^2$,label.side=right,foreground=red}{traja .. trajb}
			\fmfi{dashes,label=$\delta \phi^1$,label.side=right,foreground=red}{trajb .. trajc}
			\fmfi{dashes,foreground=blue,label=$\delta \phi^{a}$,label.side=right}{trajc .. flowb}
			\fmfi{dashes,foreground=blue,label=$\delta \phi^{b}$,label.side=right}{flowb .. flowa}
			\fmfiv{label=$\Sigma_{\rho}$,label.angle=0}{sigmac}
			\fmfiv{label=$\Sigma_{\rho'}$}{tauc}
			\fmfiv{decor.shape=circle,decor.filled=full,decor.size=0.025w,label=$y$,label.angle=-135}{flowa}
			\fmfiv{decor.shape=circle,decor.filled=full,decor.size=0.025w,label=$z$}{trajc}
			\fmfiv{decor.shape=circle,decor.filled=full,decor.size=0.025w,label=$x$,label.angle=-135}{traja}
			\fmfiv{decor.shape=circle,decor.filled=full,decor.size=0.025w,label=$x'$,label.angle=-160,label.dist=0.07w}{trajb}
			\fmfiv{decor.shape=circle,decor.filled=full,decor.size=0.025w,label=$y'$,label.angle=-90}{flowb}
		\end{fmfgraph*}}
		\hfill
		\mbox{}

		\caption{Gauge transformations in field space}
		\label{fig:gauge}
	\end{figure*}
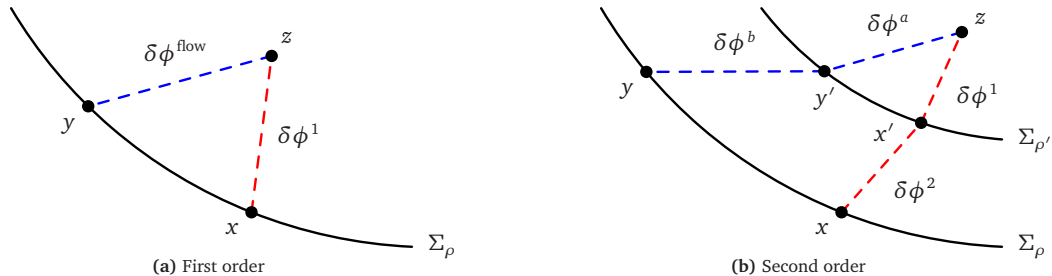
	
	\para{Linear term}%
	Consider Fig.~\ref{fig:gauge-first}.
	We wish to compute the
	coefficient $N_\alpha$
	at a field-space point $x$,
	which can be taken to lie
	on a hypersurface of fixed energy density $\rho$.
	We denote this hypersurface $\Sigma_\rho$.
	According to the separate universe approximation,
	$N_\alpha$
	can be computed
	from the number of e-folds required to flow back to
	$\Sigma_\rho$ after making a generic
	(``off-shell'')
	displacement from $x$.
	Anticipating the discussion of second-order contributions,
	we denote this displacement $\delta \phi^1$
	and write $z = x + \delta \phi^1$.
	
	The number of e-folds required to return to $\Sigma_\rho$ must be
	computed along the inflationary trajectory which passes through
	$z$. In Fig.~\ref{fig:gauge-first}, this trajectory intersects
	$\Sigma_\rho$ at $y$. The tangent to the trajectory at $y$ is the
	normal vector $\hat{n}(y)$. Therefore the (``on-shell'')
	field-space displacement along
	this trajectory, to first order in $\delta\phi^1$, is
	$\delta \phi^{\mathrm{flow}}_\alpha \approx
	- \hat{n}_\alpha \hat{n}_\beta \delta \phi^1_\beta$.
	The symbol `$\approx$' denotes equality up to higher-order
	terms in $\delta \phi^1$ which have been omitted,
	and
	we have adopted a convention in which quantities
	evaluated at $x$---such as the
	unit vector $\hat{n}$---are written without an argument.
	Combining Eqs.~\eqref{eq:field-eq} and~\eqref{eq:unit-vector},
	we conclude
	\begin{equation}
		\delta N \approx - \frac{1}{\Mp}
			\frac{\hat{n}_\alpha \delta \phi^1_\alpha}{\sqrt{2\epsilon}}
	\end{equation}
	and therefore
	\begin{equation}
		\frac{\partial N}{\partial \phi^1_\alpha}
		= - \frac{1}{\Mp}
			\frac{\hat{n}_\alpha}{\sqrt{2\epsilon}}
		= - \frac{1}{\Mp}
			\frac{\hat{n}_\alpha}{\refractiveindex} .
		\label{eq:gauge-first}
	\end{equation}
	where
	we have reintroduced the refractive index
	$\refractiveindex = \sqrt{2\epsilon}$
	defined in \S\ref{sec:slow-roll}.
	Eq.~\eqref{eq:gauge-first} is the term $N_\alpha$
	in~\eqref{eq:gauge-transformation}.
	
	\para{Quadratic term}%
	The quadratic Taylor coefficient can be obtained from
	the variation in $\partial N / \partial \phi^1_\alpha$
	under a \emph{second} generic displacement
	$\delta \phi^2$.
	Under
	this displacement the origin is shifted to $x' = x + \delta \phi^2$.
	Because
	the energy density at $x'$ will typically differ from $\rho$,
	it lies on a displaced hypersurface $\Sigma_{\rho'}$.
	However,
	the definition of
	$N$ is unchanged and
	must still be measured to the intersection with
	$\Sigma_\rho$ at $y$.
	We should compute
	the flow along the
	trajectory passing through $z$.
	The path
	$z \rightarrow y' \rightarrow y$ is a
	discrete approximation to an integral
	along this flow.
	The calculation should be carried to linear order in
	$\delta \phi^1$ and $\delta \phi^2$ independently.
	
	In Fig.~\ref{fig:gauge-second},
	the on-shell flow from $z = x' + \delta\phi^1$
	back to $\Sigma_{\rho'}$ is $\delta \phi^a$.
	Repeating the analysis above, we find
	\begin{equation}
		\delta \phi^a_\alpha \approx - \hat{n}'_\alpha \hat{n}'_\beta
		\delta \phi^1_\beta ,
		\label{eq:flow-back-a}
	\end{equation}
	where $\hat{n}'_\alpha \equiv \hat{n}_\alpha(x') \approx \hat{n}_\alpha +
	\delta \phi^2_\beta \partial_\beta \hat{n}_\alpha$.
	(It is only necessary to work to first order in $\delta \phi^2$,
	since~\eqref{eq:flow-back-a} is proportional to
	$\delta\phi^1$.)
	The on-shell flow from $y'$ back to $\Sigma_{\rho}$
	is
	\begin{equation}
		\delta \phi^b_\alpha \approx - \hat{n}''_\alpha \hat{n}''_\beta
		\Delta_\beta - \hat{n}''_\alpha
			\Big(
				\frac{K_{\beta\gamma}}{2} - \partial_\beta \hat{n}_\gamma
			\Big)
			\Delta_\beta \Delta_\gamma .
		\label{eq:displacement-b}
	\end{equation}
	We have defined $\Delta_\alpha$ to be the displacement to $y'$,
	\begin{equation}
		\Delta_\alpha \equiv \delta \phi^1_\alpha +
			\delta \phi^2_\alpha + \delta \phi^a_\alpha ,
	\end{equation}
	and $\hat{n}''_\alpha \equiv \hat{n}_\alpha(y')$.
	The symmetric tensor $K_{\alpha\beta}$ is the extrinsic curvature
	of $\Sigma_\rho$,
	or ``second fundamental form,''
	and is defined by
	$K_{\alpha \beta} \equiv h_{\alpha\gamma} h_{\beta\delta}
	\partial_\gamma \hat{n}_\delta$
	\cite{Hawking:1973uf}.
	It is related to the dilation and shear of the expansion tensor via
	\begin{equation}
		K_{\alpha\beta} = \frac{1}{\Mp \nu}
			\left(
				\frac{\theta}{d} h_{\alpha\beta} + \isosigma_{\alpha\beta}
			\right) ,
		\label{eq:extrinsic-curvature}
	\end{equation}
	where $\isosigma_{\alpha\beta}$ is the projection of the shear
	onto the isocurvature subspace,
	$\isosigma_{\alpha\beta} \equiv
	h_{\alpha\gamma} h_{\beta\delta} \sigma_{\gamma\delta}$.
	The first term in~\eqref{eq:displacement-b}
	is a linear, trigonometric approximation.
	The second is a correction for the curvature of
	$\Sigma_\rho$.
	A
	similar construction
	could be used to obtain the Taylor coefficients at any desired
	order.
	
	After computing all appropriate variations, we find
	\begin{widetext}
	\begin{equation}
		\begin{split}
		\frac{\partial^2 N}{\partial \phi^1_\alpha \partial \phi^2_\beta}
		& = -\frac{1}{\Mp}
			\left(
				\frac{K_{\alpha\beta}}{\sqrt{2\epsilon}}
				+
				\hat{n}_\alpha \partial_\beta (2\epsilon)^{-1/2}
				+
				\hat{n}_\beta \partial_\alpha (2\epsilon)^{-1/2}
				-
				\hat{n}_\alpha \hat{n}_\beta \hat{n}_\gamma \partial_\gamma
				(2\epsilon)^{-1/2}
			\right) \\
		& = -\frac{1}{\Mp \refractiveindex}
			\Big(
				K_{\alpha\beta}
				-
				\hat{n}_\alpha \isopartial_\beta \ln \refractiveindex
				-
				\hat{n}_\beta \isopartial_\alpha \ln \refractiveindex
				-
				\frac{\hat{n}_\alpha \hat{n}_\beta}{\Mp}
				\frac{\eta}{2\refractiveindex}
			\Big) ,
		\end{split}
		\label{eq:gauge-second}
	\end{equation}
	\end{widetext}
	where
	$\eta = \d \ln \epsilon / \d N$
	is the natural generalization of the single-field $\eta$-parameter.
	It measures the variation of $\epsilon$ along the adiabatic
	direction.
	To yield sufficient e-foldings, it must typically be small
	while observable scales are leaving the horizon.
	Defining
	$\adpartial \equiv \hat{n}_\alpha \partial_\alpha$
	to be a derivative along $\hat{n}_\alpha$,
	it can be written
	\begin{equation}
		\eta \equiv \frac{2\Mp}{\refractiveindex} \adpartial \epsilon .
	\end{equation}
	In addition,
	$\isopartial_\alpha \equiv h_{\alpha\beta} \partial_\beta$
	is a derivative in the plane tangent to $\Sigma_\rho$
	at $x$.
	This tangent space can be
	interpreted as the subspace of isocurvature
	modes.
	Only the $\eta$-component of~\eqref{eq:gauge-second} depends
	purely on the local behaviour of the adiabatic direction,
	and therefore the direction in field space restricted by
	the slow-roll approximation.
	The remaining terms all probe details of the isocurvature
	subspace.

	Dropping the distinction between $\delta \phi^1$ and $\delta \phi^2$,
	Eq.~\eqref{eq:gauge-second} is equal to $N_{\alpha\beta}$.
	It is
	symmetric even though we have not treated the displacements
	$\delta\phi^1$ and $\delta\phi^2$ equally.
	This is a consequence of associativity of vector addition, which makes
	$z$ the same no matter in which order we apply
	the displacements.
	The inflationary trajectory passing through $z$ is unique, so
	$N_{\alpha\beta}$ can only depend on a symmetric combination of
	$\delta\phi^1$ and $\delta\phi^2$.
	
	Eq.~\eqref{eq:gauge-second}
	shows that $N_{\alpha\beta}$
	depends on the anisotropy of $\epsilon$---or,
	in the optical interpretation, the refractive index
	$\refractiveindex$.
	It also depends on the extrinsic curvature of $\Sigma_\rho$,
	which is a function of the shape of the hypersurfaces of constant
	energy density.
	In particular, because $\hat{n}_\alpha K_{\alpha\beta} = 0$,
	this term can be interpreted as a metric on the subspace of
	isocurvature modes.
	
	\subsection{Local mode $\fNL$}
	\label{sec:local-fnl}
	
	\subpara{Two-point function}%
	These results can be combined to obtain the usual formulae for
	the amplitude of the local mode, $\fNL$.
	With our usual assumptions about the amplitude of those
	correlation functions we neglect,
	the two-point function of $\zeta$ satisfies
	\begin{equation}
	\begin{split}
		\langle \zeta(\vect{k}_1) \zeta(\vect{k}_2) \rangle
		= \mbox{} &
		(2\pi)^3 \delta(\vect{k}_1 + \vect{k}_2)
		N_{\alpha} N_{\beta}
		\Gamma_{\alpha i} \Gamma_{\beta j}
		\frac{\Sinit_{ij}}{k^3}
		\\ & \mbox{}
		+ \Or\Big(
			\frac{H^4}{\Mp^4}
		\Big) ,
	\end{split}
	\label{eq:2pf-zeta}
	\end{equation}
	where $k$ is the common amplitude of $\vect{k}_1$ and $\vect{k}_2$
	and $N_\alpha$ is the first-order component of the gauge
	transformation, Eq.~\eqref{eq:gauge-first}.
	Application of the
	chain rule to the contractions in~\eqref{eq:2pf-zeta} allows the
	Lyth--Rodr\'{\i}guez Taylor coefficients to be identified,
	\begin{equation}
		N_i \equiv \frac{\partial N}{\partial \phi_i} =
		N_\alpha \Gamma_{\alpha i} .
	\end{equation}
	It follows that~\eqref{eq:2pf-zeta} is the standard result
	\cite{Lyth:2005fi}.
	
	\subpara{Three-point function}%
	Neglecting the initial three-point function
	$\Ainit_{\kindex{i}\kindex{j}\kindex{k}}$,
	the bispectrum can be computed by similar methods.
	There is an added complication from second-order terms in the
	gauge transformation~\eqref{eq:gauge-transformation}.
	Working from~\eqref{eq:3pf-total}
	(or~\eqref{eq:3pf-shape} and~\eqref{eq:a-solution}) gives
	\begin{widetext}
	\begin{equation}
		\langle \zeta(\vect{k}_1) \zeta(\vect{k}_2) \zeta(\vect{k}_3)
		\rangle
		= (2\pi)^3 \delta(\vect{k}_1 + \vect{k}_2 + \vect{k}_3)
		(
			N_\alpha \Gamma_{\alpha mn}
			+
			N_{\alpha\beta} \Gamma_{\alpha m} \Gamma_{\beta n}
		)
		N_i N_j \Sinit_{mi} \Sinit_{nk}
		\left(
			\frac{1}{k_1^3 k_2^3}
			+ \frac{1}{k_1^3 k_3^3}
			+ \frac{1}{k_2^3 k_3^3}
		\right)
		+ \Or\left( \frac{H^6}{\Mp^6} \right) .
	\end{equation}
	We can make the identification
	\begin{equation}
	\label{eq:Nij}
		N_{ij} \equiv \frac{\partial^2 N}{\partial \phi_i \partial \phi_j}
		= N_\alpha \Gamma_{\alpha ij} +
			N_{\alpha\beta} \Gamma_{\alpha i} \Gamma_{\beta j} ,
	\end{equation}
	where $N_{\alpha\beta}$ is the second-order
	term~\eqref{eq:gauge-second}.
	The familiar approximation for the amplitude of the local mode,
	$\fNL$, follows immediately,
	\begin{equation}
	\begin{split}
		\frac{6}{5} \fNL =
			\frac{N_{mn} N_j N_k \Sinit_{mj} \Sinit_{nk}}
			{(N_q N_r \Sinit_{qr})^2}
			= \mbox{} &
			\frac{N_\alpha \Gamma_{\alpha mn} N_j N_k \Sinit_{mj} \Sinit_{nk}}
			{(N_q N_r \Sinit_{qr})^2}
			+
			\frac{N_{\alpha\beta} \Gamma_{\alpha m} \Gamma_{\beta n}
			N_j N_k \Sinit_{mj} \Sinit_{nk}}
			{(N_q N_r \Sinit_{qr})^2} .
		\\ \equiv &
		\fNL^{\phi} + \fNL^{\text{gauge}} .
	\end{split}
	\label{eq:traditional-fnl}
	\end{equation}
	\end{widetext}
	In the final step we have divided the contributions into
	an \emph{intrinsic}
	term, $\fNL^{\phi}$
	(which contains $\Gamma_{\alpha mn}$),
	and a \emph{gauge contribution} $\fNL^{\text{gauge}}$
	(which does not).
	The intrinsic term depends on the bispectrum of the
	fluctuations $\delta\phi_\alpha$.
	Eq.~\eqref{eq:gamma-2} shows that it depends on $u_{\alpha\beta\gamma}$,
	and therefore has a memory of the
	\emph{nonlinear} evolution of the connecting vectors
	along the trajectory.
	However, it has no dependence on the nonlinear part of the gauge
	transformation.
	\emph{Vice versa},
	the gauge term depends on the nonlinear part of the gauge
	transformation, and only on the \emph{linear}
	evolution of the connecting vectors---%
	that is, the Jacobi fields, in the guise of the van Vleck
	matrix~\eqref{eq:gamma-deriv}.

	This separation was first made in Ref.~\cite{Mulryne:2009kh},
	where it was shown that the gauge contribution dominated
	in a class of models known to generate large $|\fNL|$
	\cite{Byrnes:2008wi,*Byrnes:2008zy}.
	We will sharpen this division slightly in
	Eqs.~\eqref{eq:fnl-A}--\eqref{eq:fnl-B} below.
	
	The outcome of this
	discussion is that $\fNL$ could be computed efficiently
	by
	decomposing~\eqref{eq:traditional-fnl} into the component
	gauge transformations and $\Gamma$-symbols,
	which can be obtained using ordinary differential
	equations.
	An alternative
	approach is to work from the explicit formula~\eqref{eq:3pf-shape},
	yielding
	\begin{widetext}
	\begin{equation}
		\langle \zeta(\vect{k}_1) \zeta(\vect{k}_2) \zeta(\vect{k}_3)
		\rangle
		= (2\pi)^3 \delta(\vect{k}_1 + \vect{k}_2 + \vect{k}_3)
		\big(
			N_{\alpha} N_{\beta} N_{\gamma}
			\alpha_{\alpha\mid\beta\gamma}
			+
			N_{\alpha\beta} N_{\gamma} N_{\delta}
			\Sigma_{\alpha\gamma}
			\Sigma_{\beta\delta}
		\big)
		\left(
			\frac{1}{k_1^3 k_2^3}
			+ \frac{1}{k_1^3 k_3^3}
			+ \frac{1}{k_2^3 k_3^3}
		\right)
		+ \Or\left( \frac{H^6}{\Mp^6} \right) .
		\label{eq:3pf-local-form}
	\end{equation}
	\end{widetext}
	It then follows that
	\begin{equation}
		\fNL^{\phi} = \frac{5}{6}
			\frac{N_\alpha N_\beta N_\gamma \alpha_{\alpha\mid\beta\gamma}}
			{(N_\lambda N_\mu \Sigma_{\alpha\beta})^2}
			= \frac{5}{18}
			\frac{N_\alpha N_\beta N_\gamma \alpha_{\alpha\beta\gamma}}
			{(N_\lambda N_\mu \Sigma_{\alpha\beta})^2} .
 	\label{eq:fnl-intrinsic}
	\end{equation}
 	In the final equality we have defined $\alpha_{\alpha\beta\gamma}$
 	by symmetrization,
 	\begin{equation}
 		\alpha_{\alpha\beta\gamma} \equiv
 		\alpha_{\alpha\mid\beta\gamma} +
 		\alpha_{\beta\mid\alpha\gamma} +
 		\alpha_{\gamma\mid\alpha\beta} .
 	\end{equation}
 	Note that this combination is not normalized to give weight unity.
 	Eq.~\eqref{eq:alpha-flavour-transport} shows that it obeys
 	the transport equation~\eqref{eq:3pf-transport} for the three-point
 	function after dropping primes on all indices.
 	It was in this form that $\fNL$ was quoted in
 	Refs.~\cite{Mulryne:2009kh,*Mulryne:2010rp},
 	although the derivation was given in real space and is not the
 	same as the one given here.
 	
 	\para{Gauge contribution}%
	There is some interest in isolating the gauge contribution to
	$|\fNL|$. As explained above, this is known to dominate in some models,
	including examples where large $|\fNL|$ is generated during
	a turn in field space
	\cite{Byrnes:2008wi,*Byrnes:2008zy,Elliston:2011dr}.
	Comparison with~\eqref{eq:3pf-local-form} shows that it can be
	written
	\begin{equation}
		\frac{6}{5} \fNL^{\text{gauge}}
		=
		\frac{N_{\alpha\beta} N_\gamma N_\delta
			\Sigma_{\alpha\gamma} \Sigma_{\beta\delta}}
			{(N_\lambda N_\mu \Sigma_{\lambda\mu})^2} .
	\end{equation}
	
	Combining~\eqref{eq:gauge-first} and~\eqref{eq:gauge-second}
	gives an explicit expression,
	\begin{widetext}
	\begin{equation}
		\frac{6}{5} \fNL^{\text{gauge}}
		= \frac{\eta}{2}
		- \Mp \nu \left(
			\frac{\langle \sigma \delta \phi_\alpha \rangle
				K_{\alpha\beta}
				\langle \delta \phi_\beta \sigma \rangle}
				{\langle \sigma \sigma \rangle^2}
			-
			2 \frac{\langle \sigma \delta \phi_\alpha \rangle
				\isopartial_\alpha \ln \nu}
				{\langle \sigma \sigma \rangle}
		\right) ,
		\label{eq:gauge-contribution-fNL}
	\end{equation}
	\end{widetext}
	where we have defined $\langle \sigma \sigma \rangle =
	\hat{n}_\alpha \hat{n}_\beta \Sigma_{\alpha\beta}$,
	and $\langle \sigma \delta\phi_\alpha \rangle = \hat{n}_\beta
	\Sigma_{\alpha\beta}$.
	This expression is covariant under rotations
	of the isocurvature plane. However, its form suggests a natural
	coordinate basis in which its content is more transparent.
	Since $K_{\alpha\beta}$ is symmetric, it can be diagonalized.
	Its eigenvectors form an orthonormal basis
	directed along the \emph{principal curvature directions}
	of the fixed energy-density hypersurface in field space.
	We label these eigenvectors with an index $\isoindex{m}$
	and denote them
	$\zeta_{\alpha}^{\isoindex{m}}$,
	which can be considered as a vielbein.
	We can refer to the corresponding isocurvature directions
	as the \emph{principal isocurvature modes}.
	The corresponding eigenvalues of the second fundamental
	form are the \emph{principal
	curvatures} $k_\isoindex{m}$.
	
	Next, we define a correlation coefficient $\rho_{\isoindex{m}}$
	between the $\isoindex{m}^{\text{th}}$ principal isocurvature mode
	and the adiabatic direction,
	\begin{equation}
		\zeta_\alpha^{\isoindex{m}} \langle \sigma \delta\phi_\alpha \rangle
		\equiv
		\langle \sigma \isoindex{m} \rangle
		=
		\rho_{\isoindex{m}}
		\langle \sigma \sigma \rangle^{1/2}
		\langle \isoindex{m} \isoindex{m} \rangle^{1/2} .
	\end{equation}
	It is also useful to define analogues of the $\eta$-parameter for
	the isocurvature directions. It is a matter of convention how this
	is done. By analogy with our definition of $\eta$ in the
	adiabatic direction we set
	\begin{equation}
		\eta_{\isoindex{m}} \equiv
		\frac{2\Mp}{\refractiveindex}
		\zeta^{\isoindex{m}}_\alpha \isopartial_\alpha \epsilon .
	\end{equation}
	Unlike the adiabatic $\eta$-parameter, these
	isocurvature $\eta_{\isoindex{m}}$-parameters
	need not be small even if slow-roll is an excellent approximation.
	In this basis we find
	\begin{equation}
		\fNL^{\text{gauge}} =
			\frac{\eta}{2}
			+ \sum_{\isoindex{m}} \eta_{\isoindex{m}}
				 \rho_{\isoindex{m}}
				 \frac{\langle \isoindex{m} \isoindex{m} \rangle^{1/2}}
				 	{\langle \sigma \sigma \rangle^{1/2}}
			- \Mp \nu \sum_{\isoindex{m}}
				k_{\isoindex{m}} \rho_{\isoindex{m}}^2
				\frac{\langle \isoindex{m} \isoindex{m} \rangle}
					{\langle \sigma \sigma \rangle} .
		\label{eq:fnl-gauge}
	\end{equation}
	As has been explained,
	these depend only on the Jacobi fields
	and geometrical quantities
	at the time of
	evaluation for $\fNL$.
	We have not displayed the $\vect{k}$-modes associated with these
	objects.
	Eq.~\eqref{eq:fnl-gauge} strictly applies for roughly comparable
	$|\vect{k}|$.
	
	There are three contributions.
	First, there is the adiabatic $\eta$-parameter.
	As explained
	above this will almost always be negligible.
	Second, there is a weighted sum of $\eta_{\isoindex{m}}$-parameters
	associated with the isocurvature directions.
	These may be individually large.
	Their contribution is suppressed by the correlation coefficient
	$\rho_{\isoindex{m}}$
	between the adiabatic mode and
	fluctuations in the $\isoindex{m}^{\text{th}}$ direction,
	and also by the ``anisotropy factor''
	$(\langle \isoindex{m} \isoindex{m} \rangle /
	\langle \sigma \sigma \rangle)^{1/2}$
	which measures their relative amplitude.
	Third, there is a weighted sum of the principal curvatures.
	These are weighted by the combination
	$\rho_{\isoindex{m}}^2 \langle \isoindex{m} \isoindex{m} \rangle
	/ \langle \sigma \sigma \rangle$.
	Therefore, this term is typically
	dominant when the bundle has exaggerated extent in
	at least one isocurvature direction.
	
	In a two-field model, Eq.~\eqref{eq:fnl-gauge} becomes especially
	simple. There is only one principal isocurvature mode,
	and it is orthogonal to the adiabatic direction.
	Also, the second fundamental form $K_{\alpha\beta}$ has a null
	eigenvector and therefore the principal curvature $k$ is simply its
	trace. Comparison with~\eqref{eq:extrinsic-curvature}
	shows that
	\begin{equation}
		k = \tr K_{\alpha\beta} =
			\frac{1}{\Mp \nu} \left(
				\frac{M-1}{M} \theta - \hat{n}_{\alpha} \hat{n}_{\beta}
				\sigma_{\alpha\beta}
			\right) .
	\end{equation}
	As in \S\ref{sec:geo-optics}, we have set $M$ to be the dimension of
	field space.

	\subpara{Non-Gaussianity at the adiabatic limit}%
	There has been considerable interest in the fate of non-Gaussianity 
	if an adiabatic limit is reached during slow-roll inflation. 
	Meyers \& Sivanandam \cite{Meyers:2010rg,*Meyers:2011mm} 
	studied a class of models in which
	$\fNL$, $\gNL$ and $\tauNL$ decay to negligible 
	values when
	all isocurvature modes decay, and argued that this behaviour is generic. 
	However, 
	explicit examples exist in which an observable value of $\fNL$ persists 
	even after all isocurvature modes are extinguished
	\cite{Kim:2010ud,*Kim:2011jea,Elliston:2011dr,Mulryne:2011ni}. 
	The separation of $\fNL$ into intrinsic and gauge contributions 
	allows us to shed further light on this issue. 

	At an adiabatic limit we expect $\langle \isoindex{m} \isoindex{m}
	\rangle \rightarrow 0$,
	and therefore
	Eq.~\eqref{eq:fnl-gauge} implies $\fNL^{\text{gauge}} \approx \eta/2$.
	The same conclusion can be obtained from~\eqref{eq:gauge-contribution-fNL}
	because any tensor projected onto the isocurvature plane
	(such as $K_{\alpha\beta}$ or $\isopartial_\alpha$)
	is orthogonal to $\Sigma_{\alpha\beta}$
	in this limit.
	This is an advantage of the tensorial approach
	we have described,
	based on associating isocurvature modes
	with the tangent plane to surfaces
	of constant energy density in phase space.
	
	One can also show that the intrinsic $\fNL$ satisfies
	\begin{equation}
		\fNL^{\phi}
		= \fNL^{\phi,\adlimit}
		+ \frac{\eta_{\adlimit}}{2}
		- \frac{\eta}{2} ,
		\label{eq:fnl-phi}
	\end{equation}
	where `$\adlimit$' denotes evaluation just after the adiabatic limit
	is reached.
	In the language of \S\ref{sec:slow-roll-focusing}
	this may coincide with the onset of an inflow trajectory.
	We conclude that, at any subsequent time, $\fNL$ has value
	\begin{equation}
		\fNL = \fNL^{\phi,\adlimit} + \frac{\eta_{\adlimit}}{2} ,
		\label{eq:fnl-adiabatic-limit}
	\end{equation}
	which is constant as we expect.
	If the adiabatic limit is reached during slow-roll inflation,
	where $\eta_{\adlimit}$ must be small,
	this enables us to give a more precise formulation
	of Meyers \& Sivanandam's argument:
	if $\fNL$ is large in the adiabatic limit, it
	must be because a large \emph{intrinsic} three-point function
	is developed during the evolution.
	This is indeed the case in known examples where a large
	$\fNL$ is reached in the ``horizon-crossing approximation''
	\cite{Kim:2006ys,Kim:2010ud,Elliston:2011dr}.
	
	Eqs.~\eqref{eq:fnl-gauge}, \eqref{eq:fnl-phi}
	and~\eqref{eq:fnl-adiabatic-limit} also enable us to sharpen the
	division between ``gauge'' and ``intrinsic'' contributions. We define
	\begin{subequations}
	\begin{align}
		\fNL^A & = \fNL^\phi + \frac{\eta}{2}
		\label{eq:fnl-A}
		\\
		\fNL^B & = \fNL^{\mathrm{gauge}} - \frac{\eta}{2} .
		\label{eq:fnl-B}
	\end{align}
	\end{subequations}
	The advantage of this redefinition is that
	the $A$- and $B$-type contributions are constant at an adiabatic limit;
	indeed, $\fNL^B$ is zero there
	because it captures only transient
	effects caused by the evolving isocurvature modes.
	However, when $|\fNL|$ is large
	the $A$- and $B$-type terms approximately correspond to
	the intrinsic and gauge $\fNL$.
	
	This division is not unique, because a total derivative can always be
	added to the time integral in $\Gamma_{\alpha ij}$.
	However, the division
	in Eqs.~\eqref{eq:fnl-A}--\eqref{eq:fnl-B}
	seems phenomenologically useful
	because all models (of which we are aware) which generate large
	non-gaussianity do so in one of two ways:
	\emph{either} $\fNL^A$ becomes large at the adiabatic limit,
	\emph{or} $\fNL^B$ is large some time before the adiabatic limit
	is reached.
	As the following examples show,
	the underlying reason seems to be that
	the $B$-type term responds immediately to
	strong distortions of the shape of bundle, whereas the $A$-type
	term does not.
	
	\subpara{Example: Byrnes~{\etal} model}%
	We illustrate Eqs.~\eqref{eq:gauge-contribution-fNL},
	\eqref{eq:fnl-gauge}
	and~\eqref{eq:fnl-A}--\eqref{eq:fnl-B}
	using examples drawn from
	the literature.
	
	Consider the model
	$V = V_0 \phi^2 \e{-\lambda \chi^2}$
	introduced by Byrnes~{\etal}~\cite{Byrnes:2008wi}.
	We follow their choices,
	setting $\lambda = 0.05 \Mp^{-2}$
	and fixing initial conditions
	$\phi = 16 \Mp$ and $\chi = 0.001\Mp$.
	The first phase of evolution is descent from a ridge, during
	which a large spike in $\fNL$ is generated
	by the gauge term.
	An interpretation of this contribution was given in
	Ref.~\cite{Elliston:2011dr}.
	
\begin{figure*}
		\hfill
		\subfloat[][$\fNL$\label{fig:byrnes-fNL}]{
			\includegraphics[scale=0.6]{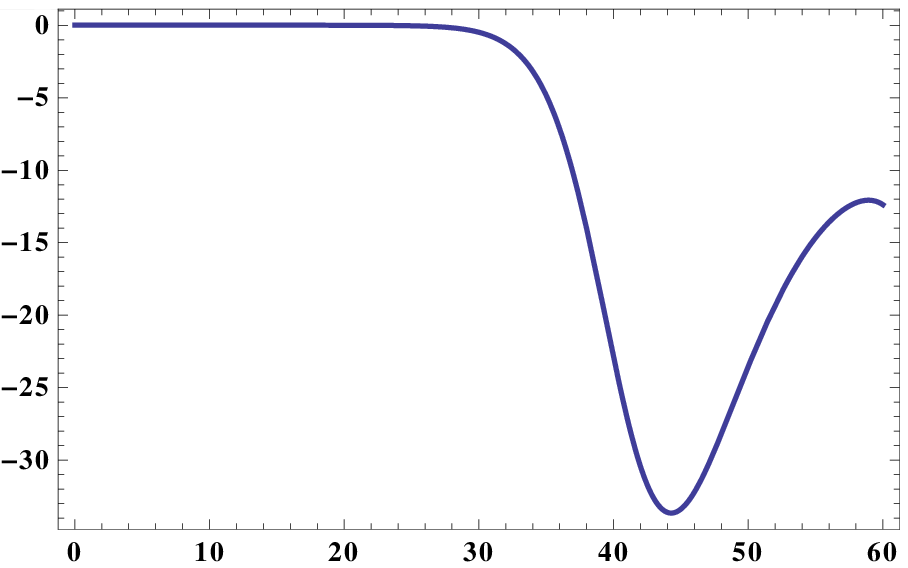}
		}
		\hfill
		\subfloat[][difference between full $\fNL$ and
			curvature contribution\label{fig:byrnes-fNLdiff}]{
			\includegraphics[scale=0.6]{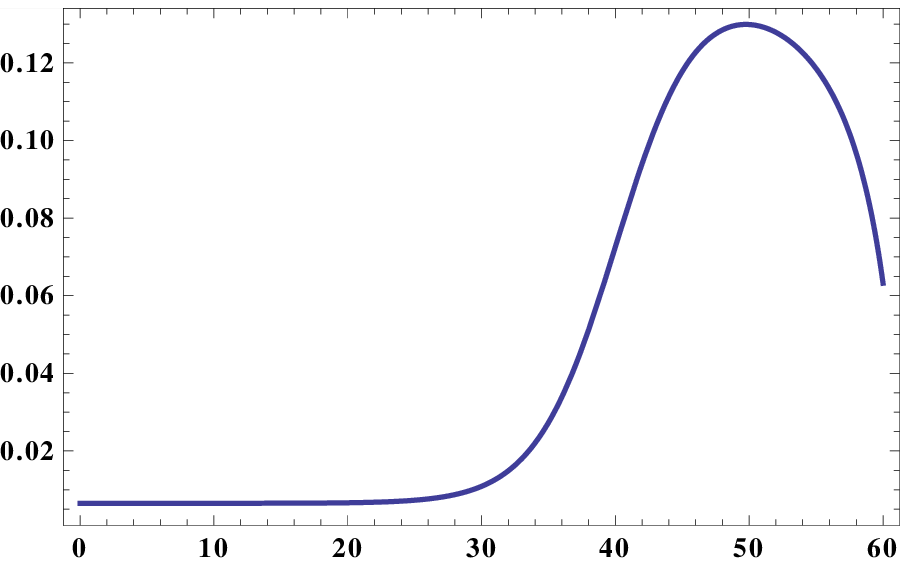}
		}
		\hfill
		\mbox{}
		\\
		\hfill
		\subfloat[][anisotropy factor $(\langle \isoindex{m}
			\isoindex{m} \rangle / \langle \sigma \sigma \rangle)^{1/2}$
			\label{fig:byrnes-anisotropy}]{
			\includegraphics[scale=0.6]{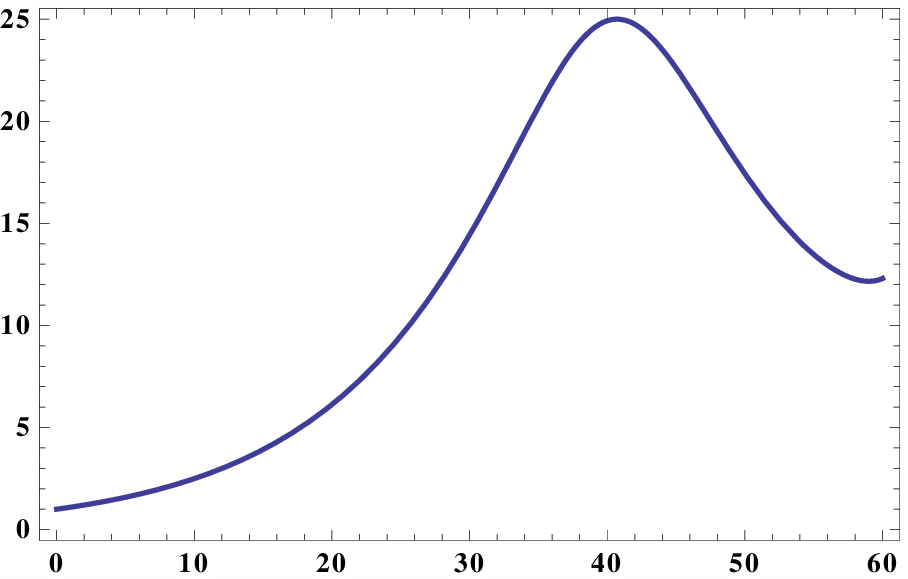}
		}
		\hfill
		\subfloat[][correlation coefficient $\rho_{\isoindex{m}}$
			\label{fig:byrnes-correlation}]{
			\includegraphics[scale=0.6]{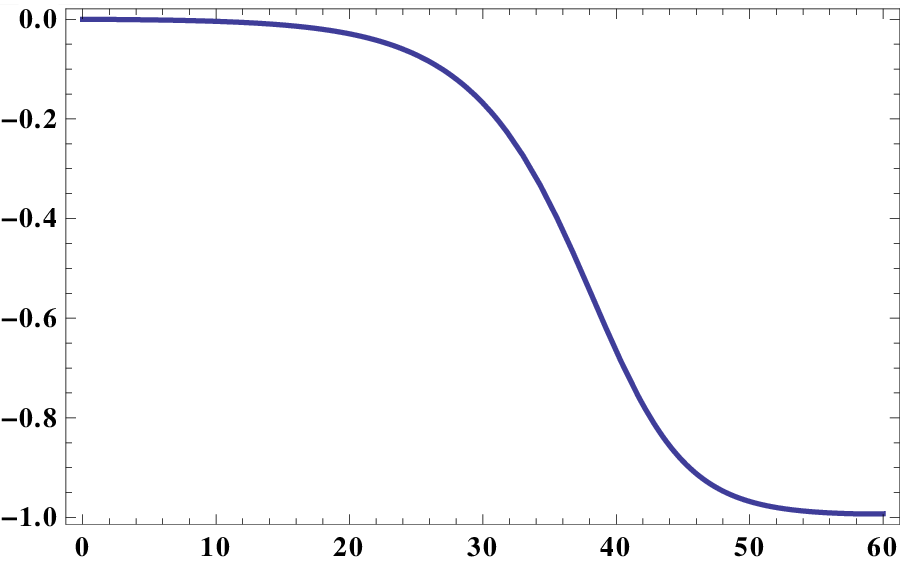}
		}
		\hfill
		\mbox{}
		\\
		\hfill
		\subfloat[][principal curvature $k_{\isoindex{m}}$
			\label{fig:byrnes-curvature}]{
			\includegraphics[scale=0.6]{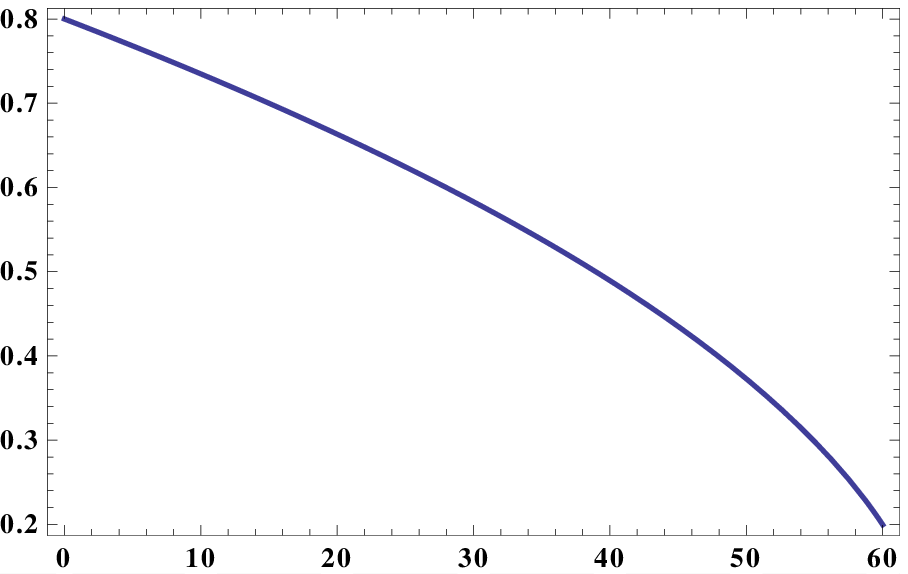}
		}
		\hfill
		\subfloat[][isocurvature $\eta_{\isoindex{m}}$
			\label{fig:byrnes-eta}]{
			\includegraphics[scale=0.6]{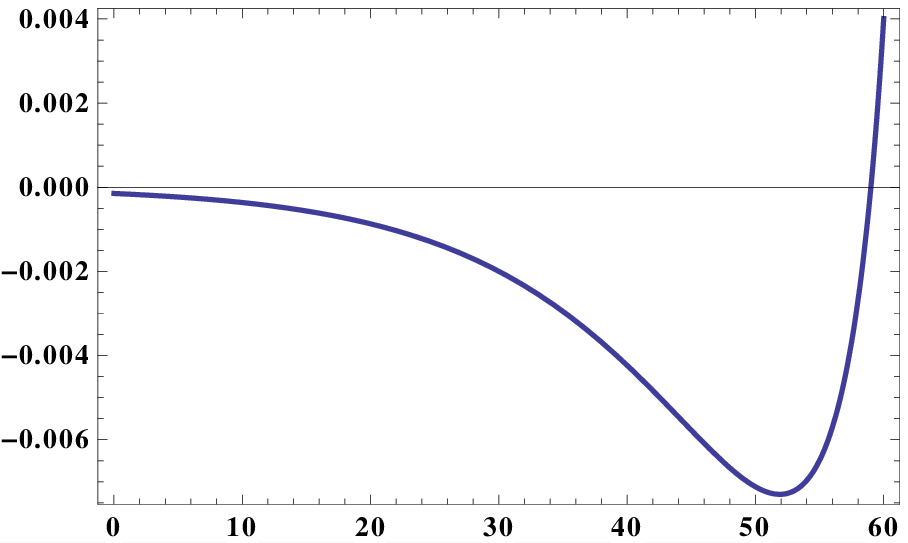}
		}
		\hfill
		\mbox{}
		\\
		\hfill
		\subfloat[][dilation $\theta$
			\label{fig:byrnes-dilation}]{
			\includegraphics[scale=0.6]{Diagrams/Byrnes/curvature}
		}
		\hfill
		\subfloat[][focusing $\exp \int \theta \, \d N$
			\label{fig:byrnes-focusing}]{
			\includegraphics[scale=0.6]{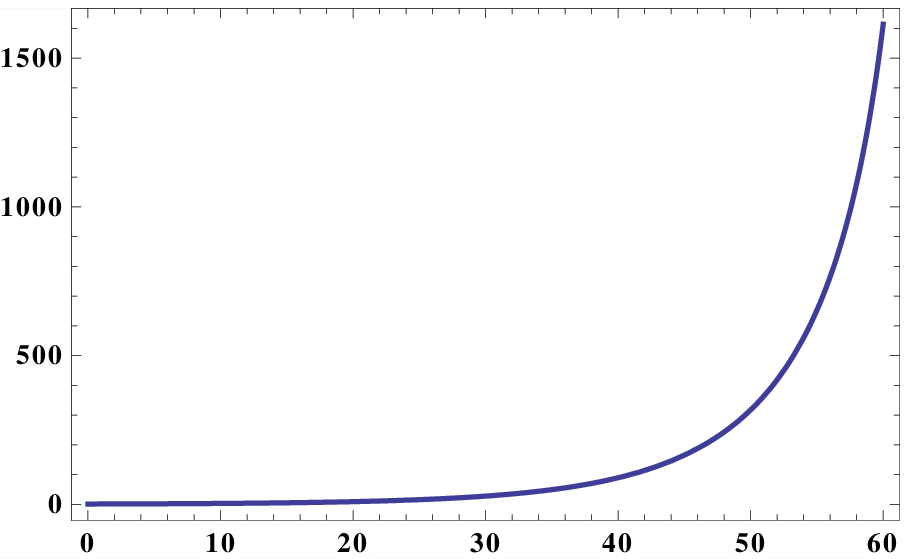}
		}
		\hfill
		\mbox{}
		\caption{Bundle parameters for the
		Byrnes~{\etal} model
		$V = V_0 \phi^2 \e{-\lambda \chi^2}$.
		The initial conditions are $\phi = 16\Mp$
		and $\chi = 0.001\Mp$,
		and $\lambda = 0.05 \Mp^{-2}$.
		All plots are against the e-folding number $N$,
		measured from horizon exit of the mode in question.
		\label{fig:bundle-byrnes}}
	\end{figure*}

	In Fig.~\ref{fig:byrnes-fNL} we plot $\fNL$
	during the inflationary phase.
	For most of the evolution it is dominated by
	$\fNL^{\text{gauge}}$.
	In turn $\fNL^{\text{gauge}}$ is dominated by the extrinsic
	curvature term $K_{\alpha\beta}$.
	In Fig.~\ref{fig:byrnes-fNLdiff} we plot the difference between
	the full $\fNL$ and the $K_{\alpha\beta}$-term, demonstrating
	explicitly that it is small.
	
	In Figs.~\ref{fig:byrnes-anisotropy}--\ref{fig:byrnes-eta}
	we plot the bundle parameters which determine
	the $K_{\alpha\beta}$-term and the other contributions to
	$\fNL^{\text{gauge}}$.
	The correlation constant is initially zero
	but approaches $-1$, making the curvature and isocurvature
	mode (anti-) correlated,
	as first discussed by Langlois~\cite{Langlois:1999dw}.
	The principal curvature $k$ and isocurvature $\eta$-parameter
	exhibit only modest evolution over the entire range of
	e-folds.
	In comparison, the anisotropy factor
	$(\langle \isoindex{m} \isoindex{m} \rangle /
	\langle \sigma \sigma \rangle)^{1/2}$ grows dramatically.
	Its evolution is the dominant factor which determines the
	evolution of $\fNL$.
	A large $\fNL$
	arises because the ensemble
	of separate universes becomes highly anisotropic,
	with nearly twenty-five times as much power
	in the isocurvature direction as in the adiabatic direction.
	Evidently this must arise from a large contribution to the
	integrated shear in the propagator matrix.

	Note that, although $\fNL^B \approx \fNL^{\mathrm{gauge}}$
	responds immediately to this strong
	anisotropy factor, there is no
	corresponding significant enhancement of the
	intrinsic three-point function.
	
	In Fig.~\ref{fig:byrnes-dilation}
	and~\ref{fig:byrnes-focusing} we plot the bundle dilation,
	$\theta$, and the focusing $\Theta$.
	The dilation is always positive,
	so the bundle cross-section grows monotonically.
	Hence the total power in the isocurvature mode also grows
	monotonically.
	Evidently,
	the spike in $\fNL$ is not due to the \emph{total}
	isocurvature power, but to its relative growth compared with
	the adiabatic power.
	The large $\Theta$ implies that this model does not reach
	an adiabatic limit,
	and some other mechanism must be invoked to end
	inflation and determine the value of each observable.
	In Ref.~\cite{Byrnes:2008wi} it was assumed that sudden
	destabilization of a waterfall field could play this role.
	
\begin{figure*}
		\hfill
		\subfloat[][$\fNL$
			\label{fig:aq-fNL}]{
			\includegraphics[scale=0.6]{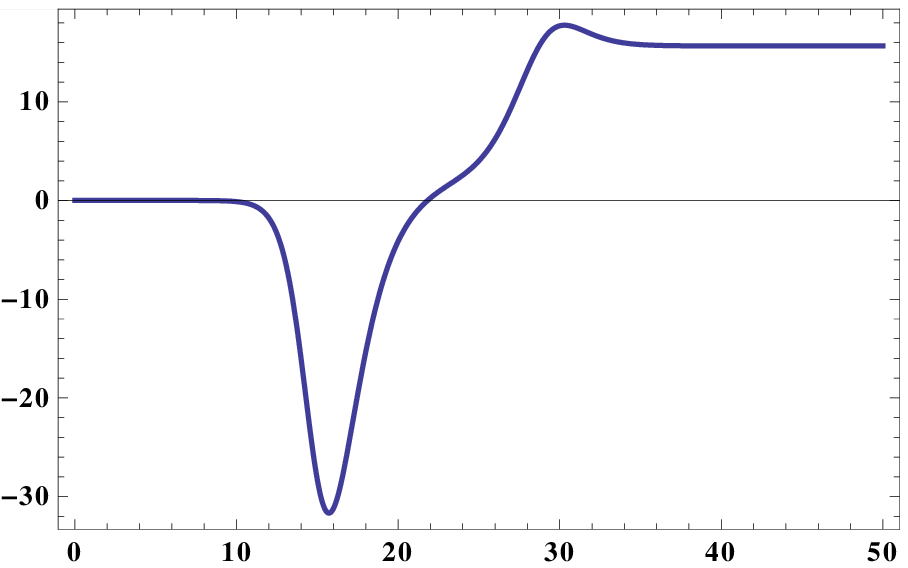}
		}
		\hfill
		\subfloat[][$\fNL^{\phi}$ (blue curve)
			and $\fNL^{\text{gauge}}$ (red curve)
			\label{fig:aq-fNLs}]{
			\includegraphics[scale=0.6]{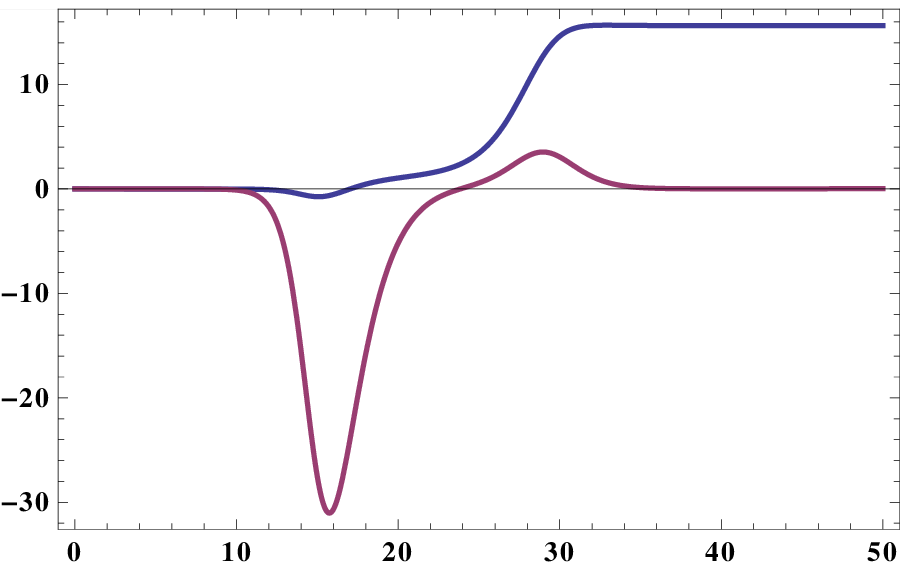}
		}
		\hfill
		\subfloat[][anisotropy factor $(\langle \isoindex{m}
			\isoindex{m} \rangle / \langle \sigma \sigma \rangle)^{1/2}$
			\label{fig:aq-anisotropy}]{
			\includegraphics[scale=0.6]{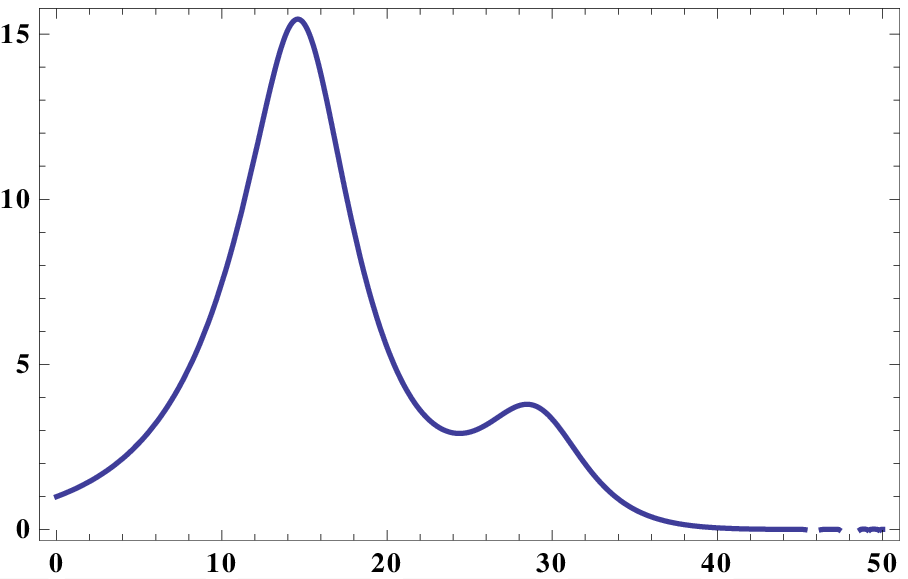}
		}
		\hfill
		\mbox{}
		\caption{Bundle parameters for the
		axion--quadratic model;
		for the potential and initial conditions,
		see Fig.~\ref{fig:focusing-aq}.
		All plots are against the e-folding number $N$,
		measured from horizon exit of the mode in question.
		\label{fig:bundle-aq}}
	\end{figure*}

	\subpara{Example: axion quadratic model}%
	A similar phenomenon occurs in the axion--quadratic model
	discussed above.
	We plot the evolution of $\fNL$ in
	Fig.~\ref{fig:aq-fNL}.
	It exhibits
	three distinct components.
	The first is a negative spike, generated
	by the axion rolling off its hilltop.
	The second is a smaller positive spike produced by
	the axion rolling into its minimum.
	These two spikes come from the gauge contribution to
	$\fNL$,
	as clearly shown in Fig.~\ref{fig:aq-fNLs}.
	Fig.~\ref{fig:aq-anisotropy}
	shows that each
	spike is inherited from a spike in the anisotropy
	factor.
	This is consistent with the analysis of
	Elliston~{\etal}~\cite{Elliston:2011dr}, in which the spikes
	were interpreted as due
	to strong deformations in the shape of the bundle.
	In the present interpretation,
	the differing signs arise because the principal
	curvature changes sign in the intermediate evolution.
	
	As for the Byrnes~{\etal} model, the
	intrinsic term $\fNL^A \approx \fNL^\phi$ does not
	respond immediately to this strong anisotropy,
	growing only later
	on approach to the adiabatic limit.
	The anisotropy is due to
	a strong shearing effect
	arising near the turn from dominantly $\phi$-evolution
	to dominantly $\chi$-evolution.
	Near the deep negative spike in $\fNL$, there is an enhancement
	in the shear oriented parallel to the principal isocurvature
	mode.
	This enhances the fluctuations in the isocurvature direction.

	The third feature is the flat plateau
	at late times, associated with the adiabatic limit.
	Fig.~\ref{fig:aq-fNLs} shows that this comes from
	growth in the intrinsic term $\fNL^{\phi}$;
	see the discussion in
	Refs.~\cite{Kim:2010ud,*Kim:2011jea,Elliston:2011dr}.

	\section{Summary}
	\label{sec:summary}
	
	In this paper we have developed an analogy between inflationary
	perturbation theory and geometrical optics. Here we
	summarize the main steps in the discussion.
	
	\para{Background}%
	In inflationary perturbation theory, we are interested in
	following the statistical properties---as measured by the
	correlation functions---of an ensemble of spacetime regions.
	This ensemble can be constructed equally well within the
	separate universe picture or traditional cosmological perturbation
	theory.
	
	The ensemble picks out a cloud of points in phase space.
	In the limit $k/aH \ll 0$, interactions between members of the
	ensemble are suppressed and each point moves along a
	phase space orbit of
	the unperturbed system.
	Therefore the ensemble traces out a narrowly-collimated
	``spray'' or bundle of trajectories.
	Where slow-roll applies, the momenta are determined in terms of
	the fields and we can work in terms of a simplified flow
	on field space.
	
	\para{Optical quantities}%
	Geometrical properties of the bundle of trajectories
	can be used to describe its evolution
	and determine its
	statistical properties.
	The quantities of principal importance
	are obtained by decomposing the expansion tensor, yielding
	the dilation, shear and twist.
	These are well-known from the description of light rays
	in general relativity.
	
	\para{Jacobi fields and van Vleck matrix}%
	The dilation, shear and twist
	determine the evolution of \emph{Jacobi fields},
	which describe infinitesimal vectors connecting nearby
	trajectories.
	At any point in the flow,
	the van Vleck matrix aggregates the linearly independent
	Jacobi fields.
	The Jacobi fields themselves are measured from a fiducial
	trajectory, which can be thought of as the eikonal of
	geometrical optics. This analogy is exact
	within the slow-roll approximation.
	
	We have argued that different implementations of the separate universe
	assumption---such as the Lyth--Rodr\'{\i}guez Taylor expansion, or
	the transport equations of~\S\ref{sec:transport}---can
	be thought of as different methods to compute the Jacobi fields,
	in the form of the van Vleck matrix~\eqref{eq:gamma-deriv}.
	More generally, the same is true for all approaches to perturbation
	theory in the limit $k/aH \rightarrow 0$.
	The most familiar implementation of the separate universe
	assumption, the ``$\delta N$ formalism'' or
	Taylor expansion approach, follows from Jacobi's method of
	varying a solution with respect to its constants of integration.
	Conversely, the transport equations arise more naturally
	from Jacobi's differential equation.
	
	On approach to a caustic, some number of Jacobi fields decay.
	At an \emph{adiabatic caustic},
	defined in~\S\ref{sec:adiabatic},
	all but one of the Jacobi fields decay.
	The single remaining field represents fluctuations
	along the caustic.
	In inflation this mode is the adiabatic fluctuation.
	The other Jacobi fields represent isocurvature
	fluctuations between the spacetime regions which make up the
	ensemble.
	Therefore, focusing
	at an adiabatic caustic
	can be interpreted as decay of isocurvature modes,
	or approach to an adiabatic limit in the sense of
	Elliston~{\etal}~\cite{Elliston:2011dr}.
	
	\para{Transport equations}%
	The ``$u$-tensors''
	encode evolution of the connecting vector fields.
	We have argued that these tensors can be computed using
	either cosmological perturbation theory or the separate universe
	approximation.
	More generally, any formalism which can reproduce the $\vect{k}$-space
	deviation equation~\eqref{eq:k-deviation}
	will reproduce the correct correlation functions,
	because
	the $u$-tensors uniquely determine the
	transport equations.
	Therefore the $u$-tensors may be used as an objective way
	to compare competing formalisms.

	The
	transport equations obtained in this way
	are generalizations of the transport equations previously
	introduced in Ref.~\cite{Mulryne:2009kh,*Mulryne:2010rp}.
	Because they are expressed in terms of $u$-tensors,
	it follows that
	they can be integrated in terms of the
	Jacobi fields and their derivatives.
	Therefore the correlation functions
	can be expressed using the van Vleck matrix
	and its derivatives.
	(Technically it is the inverse of the van Vleck matrix
	which appears, in the form of the propagator
	matrix~\eqref{eq:propagator}.)

	In turn the van Vleck matrix can be expressed in terms of
	the integrated dilation, shear and twist.
	This makes it possible to diagnose regions where the flow may
	become adiabatic by tracking the behaviour of the focusing parameter
	$\Theta$, defined in Eq.~\eqref{eq:focusing-def},
	and the behaviour of the shear and twist.

	Working within the slow-roll approximation
	we have argued that $\Theta \gtrsim 1$ implies
	the presence of remaining
	isocurvature modes.
	To be compatible with experiment,
	these must almost certainly decay before the
	surface of last scattering.
	The consequent transfer of power into the adiabatic
	mode can
	change the value of $\zeta$.
	
	\para{Flow equations}%
	The Jacobi fields yield a formal solution for each
	correlation function, analogous to the ``line of sight''
	used to simplify integration of the Boltzmann equation in
	CMB codes.
	
	This formal solution demonstrates explicitly that the transport
	equations reproduce the Taylor expansion algorithm of
	Lyth \& Rodr\'{\i}guez.
	In doing so we also obtain explicit expressions for the
	Taylor coefficients
	$\Gamma_{\alpha i}$
	and
	$\Gamma_{\alpha ij}$
	in terms of integrals of
	the expansion tensor and its derivatives along the flow.
	Similar expressions had previously been obtained by
	Yokoyama~{\etal}~\cite{Yokoyama:2007uu, *Yokoyama:2007dw}.
	
	These explicit expressions can be manipulated to obtain
	a closed set of evolution equations
	for the Taylor coefficients.
	These are
	Eqs.~\eqref{eq:k-propagator}
	and~\eqref{eq:gamma2-evolve}.
	Such equations are extremely helpful in practice, because it means the
	Taylor coefficients can be obtained without
	the challenging problem of extracting a variational derivative
	after numerical integration: without a sufficiently
	accurate integration algorithm, the small variation of
	interest can be swamped by numerical noise.
	
	\para{Transport of shape coefficients}%
	Even after obtaining the Taylor coefficients, it is necessary
	to extract coefficients for each type of momentum dependence
	(or ``shape,'' in inflationary terminology)
	which occurs in a correlation function.
	An alternative is to return to the full $\vect{k}$-space
	transport equations and derive evolution equations for these
	coefficients directly.
	The first nontrivial case is the three-point function,
	whose shape coefficients are determined by
	Eq.~\eqref{eq:alpha-flavour-transport}.
	
	\para{Gauge transformations}%
	Specializing to the slow-roll approximation, where the flow
	can be described in field space, ray-tracing techniques
	can be used to obtain the gauge transformation to $\zeta$.
	In this way the gauge transformation is expressed using geometrical
	quantities in field space, rather than merely derivatives of
	the potential.
	
	In models where a large $\fNL$ is obtained from the gauge
	transformation, this gives a geometrical interpretation of its
	magnitude. The contributory factors are:
	(1) the $\eta$-parameters of the adiabatic and principal
	isocurvature modes;
	(2) the principal curvatures of
	uniform-density hypersurfaces in field space;
	(3) the correlation coefficient
	between the adiabatic fluctuations and
	the fluctuations in each principal isocurvature mode;
	and (4) an anisotropy factor which measures
	distortions in the cloud of field-space points representing
	the ensemble.
	
	In two cases where a large, transient contribution to $\fNL$ has
	been observed, we show this principally arises from
	a strong enhancement in the anisotropy factor.

	\para{Comparison with other geometrical formulations}%
	In common with all other approaches to the evolution of
	correlation functions,
	the interpretation described in \S\S\ref{sec:transport}--\ref{sec:gauge}
	is a reformulation of perturbation theory.
	All approaches carry the same physical content.
	Therefore, aside from practical considerations,
	the merit of each reformulation arises
	from the insight gained by emphasis on different
	structures.

	The formulation we have given emphasizes the
	background phase space manifold, which encodes the structure of the
	theory in its geometry.
	This geometrical structure is mapped out by the behaviour
	of the trajectories flowing over it.
	Globally, this connection is
	made precise by the methods of Morse theory.
	Locally, it is encoded in the Jacobi fields whose role we have
	highlighted.
	
	Attempts to reformulate perturbation
	theory in terms of geometrical objects have already attracted
	attention by various authors.
	Gordon {\etal}~\cite{Gordon:2000hv}
	and
	Nibbelink \& van Tent \cite{GrootNibbelink:2001qt}
	formulated perturbation theory for the two-point function
	in terms of the Frenet basis, which they called the
	``kinematical basis.''
	(See also Ach\'{u}rcarro~{\etal}~\cite{Achucarro:2010da}.)
	Peterson \& Tegmark later extended this approach to the
	three-point function
	\cite{Peterson:2010np,*Peterson:2010mv,*Peterson:2011yt}.
	A Frenet basis can be defined for each trajectory,
	and the Frenet--Serret equation describes how this basis is transported
	along the trajectory.
	In Refs.~\cite{Gordon:2000hv,GrootNibbelink:2001qt,Peterson:2010np,
	*Peterson:2010mv,*Peterson:2011yt,Achucarro:2010da}
	these equations are used to describe transfer
	between the adiabatic and isocurvature modes.
	
	In the Frenet formulation, the isocurvature modes are identified
	with the normal, binormal, {\ldots},
	vectors.
	In our formulation
	these modes arise from the eigenvectors of the
	extrinsic curvature, $K_{\alpha\beta}$,
	which we have described as the principal isocurvature modes.
	The tangent plane spanned by the Frenet normal, binormal, {\ldots},
	is the same as the subspace spanned by the eigenvalues
	of $K_{\alpha\beta}$, so
	the physical content of these formulations is the same.
	More generally, in our formulation the properties of the
	isocurvature modes are expressed using the familiar mathematical
	apparatus used to describe hypersurfaces---normal vectors,
	first and second fundamental forms, and so on.
	
	In addition,
	we explicitly separate a ``local'' contribution
	to each $\zeta$ correlation function, arising
	from a gauge transformation and depending on
	the precise orientation of the Frenet basis,
	from the ``integrated'' contributions,
	obtained by solving the transport equations.
	Although it is clear that one can
	equally well express the integrated contributions
	in
	\emph{any} suitable basis,
	it requires extra effort to rotate
	to the Frenet basis at each step in the integration.
	We feel it is preferable to express the evolution equations of
	perturbation theory in terms of the
	original basis on field space.
	
	\para{Future directions}%
	This formalism can be extended in several directions.
	
	First,
	at some points in the discussion we specialized to the slow-roll
	approximation, to take advantage
	of certain simplifications---such as
	the twist-free and hypersurface-orthogonal character of flow.
	However, as we have presented it, the underlying formalism is
	independent of slow-roll.
	It can be used to
	evolve both field and momentum perturbations.
	This is desirable because future data from
	microwave background or galaxy surveys will be highly accurate,
	demanding commensurate accuracy in our theoretical calculations.
	
	Second,
	in this paper we have
	interpreted the decay of isocurvature modes,
	and approach to an adiabatic limit,
	as focusing of the bundle to an ``adiabatic'' caustic.
	Our detailed discussion was restricted to field space.
	It should also be possible to study focusing and decay of
	isocurvature modes on the full phase space,
	providing a framework for the study of \emph{kinetically}
	dominated scenarios, such as descent through the waterfall of
	hybrid inflation, where focusing may also occur.
	
	Third,
	the existence of explicit expressions
	for the Taylor coefficients $\Gamma_{\alpha i}$
	and $\Gamma_{\alpha ij}$
	may enable new analytic solutions to be found.
	
	Finally,
	the entire formalism can be extended to
	higher $n$-point functions.
	The case of principal interest is
	the four-point function.
	In contrast to the three-point function, this requires
	two shape parameters which determine
	$\tauNL$ and $\gNL$.

	\begin{acknowledgments}
		DS was supported by the Science and Technology Facilities Council
		[grant numbers ST/F002858/1 and ST/I000976/1],
		and would like to thank the Harish-Chandra Research Institute,
		Allahabad, for their hospitality during the programme
		\emph{Primordial Features and Non-Gaussianities} (December 2010)
		where an early version of this work was presented.
		JF was supported by the Science and Technology Facilities Council
		[grant number ST/1506029/1].
		DJM was supported by the Science and Technology Facilities Council
		[grant number ST/H002855/1].
		RHR was supported by Funda\c{c}\~{a}o para a Ci\^{e}ncia e a Tecnologia 
		through the grant SFRH/BD/35984/2007,
		and acknowledges a research studentship from the
		Cambridge Philosophical Society.
		We would like to thank Antony Lewis and Andrew Liddle
		for comments on a draft version of the manuscript.
	\end{acknowledgments}

	\appendix
	
	\section{Yokoyama~{\etal} backwards formalism}
	\label{appendix:backwards}
	
	In \S\ref{sec:flow-deltaN} we presented integral formulae for 
	the ``$\delta N$''
	coefficients, Eqs.~\eqref{eq:dn1}--\eqref{eq:dn2},
	and noted that 	
	essentially 
	identical expressions had been presented by
	Yokoyama~{\etal}~\cite{Yokoyama:2007uu,*Yokoyama:2007dw,*Yokoyama:2008by}.
	(However, Yokoyama~{\etal} obtained their results
	by very different means.)
	Since their work is closely related
	to our own in content and 
	outlook, we take this opportunity to review and extend their results.

	Their aim is to develop evolution
	or ``transport'' equations (in our terminology)
	for
	objects closely related to observation, such as the derivatives 
	$N_i$---defined in section \S\ref{sec:gauge}---%
	and the $\fNL$ parameter. They proceed as we do, first 
	fixing a flat initial hypersurface.
	In our notation this is distinguished 
	with lower case Roman indices.
	Unlike us, they also fix 
	the final slice to be the precise time at which we wish to
	know the value of each observable quantity.
	In our notation this slice is labelled with Greek letters,
	and we obtain the matrices
	$\Gamma_{\alpha i}$ and $\Gamma_{\alpha ij}$ as a function of it.
	Observables can be obtained after evaluating these functions
	at the time of interest.
	Instead,
	Yokoyama~{\etal} 
	consider intermediate flat slices \emph{between} the initial and 
	final slices,
	and express their answers as a function of the intermediate time.
	As we now explain,
	observables are to be obtained by setting
	this intermediate slice equal to the \emph{initial} hypersurface.

	For convenience we extend our index notation,
	and label quantities evaluated on the intermediate
	slice with upper case Roman 
	indices.
	Yokoyama~{\etal} introduce the quantity
	\begin{equation}
	\label{eq:NderivateInt} 	
		N_I = N_\alpha \Gamma_{\alpha I} ,
	\end{equation}
	where $N_\alpha$ was defined in \S\ref{sec:gauge}. 	
	$N_I$ is the derivative of the number of e-folds between 
	an intermediate flat hypersurface and the final uniform density 	
	hypersurface, with respect to the field values on the intermediate 
	slicing. Yokoyama~{\etal} introduce a further quantity 
	$\Theta_I = \Gamma_{I i} N_i$.
	Note that $\Theta_I$ is not to be confused with the focusing
	parameter $\Theta$ defined in the text, which is the exponential
	of the integrated dilation.
	$N_I$ and $\Theta_I$ obey the autonomous transport equations
	\begin{eqnarray}
	\frac{\d N_I}{\d N} &=& -u_{JI} N_J, \label{eq:NI}\\
 	\frac{\d \Theta_I}{\d N} &=& u_{IJ} \Theta_J, \label{eq:TI}
	\end{eqnarray}
	Evaluating $N_I$ at the final hypersurface gives $N_\alpha$,
	which provides a boundary condition for the differential equation.
	One can then evolve \emph{backwards} in time
	until we reach the initial slice.
	At this point $N_I$ will equal $N_i$, which is the Taylor coefficient
	we set out to calculate.
	After this has been done,
	$\Theta_I$ can be evolved
	\emph{forwards} from the initial hypersurface
	with boundary condition
	$\Theta_i = N_i$.
	
	We describe this as the ``backwards'' formalism, to be
	contrasted with the ``forwards'' formalism we have described in the text.
	
	The introduction of these quantities is ingenious.
	Employing
	Eq.~\eqref{eq:Nij} together with Eq.~\eqref{eq:dn2} yields
	\begin{equation}
	N_{ij} = N_\alpha \Gamma_{\alpha l} \int^N_{N*} \Gamma^{-1}_{l \sigma} 
	u_{\sigma \beta \gamma} \Gamma_{\beta i} \Gamma_{\gamma j}dN' +
	N_{\alpha \beta} \Gamma_{\alpha i} \Gamma_{\beta j},
	\end{equation}
	In turn this leads to
	\begin{equation}
	N_iN_{ij}N_j = \int^N_{N*} N_I u_{IJK} \Theta_J \Theta_k dN' + 	
	\Theta_\alpha \Theta_\beta N_{\alpha \beta} .
	\end{equation}
	Therefore, $\fNL$ can be evaluated with knowledge only of
	$N_I$, $\Theta_I$ and $u_{IJK}$.
	
	In performing this calculation, Yokoyama~{\etal}
	traded a three-index object (either
	$\Gamma_{\alpha ij}$ or $\alpha_{\alpha|\beta\gamma}$,
	depending which formulation
	is in use) for
	two one-index objects, $N_I$ and $\Theta_I$.
	This involves fewer equations and therefore can be numerically
	advantageous.
	
	Nevertheless, the backwards formalism
	has some disadvantages.
	First, because it computes only the Taylor coefficients,
	information about isocurvature modes is discarded.
	The evolution equations for
	$\Sigma_{\alpha\beta}$
	and $\alpha_{\alpha|\beta\gamma}$,
	or
	$\Gamma_{\alpha i}$ and $\Gamma_{\alpha ij}$,
	allow the isocurvature modes to be retained.
	
	Second, to obtain information about the time-evolution of
	any observable it is necessary to recalculate $N_I$ and $\Theta_I$ with
	multiple \emph{final} times.
	Although the method yields $N_I$, which is apparently related
	to the gauge transformation at an intermediate time,
	this is not quite correct.
	$N_I$ is defined for a fixed \emph{future} rather than
	past boundary condition,
	and therefore gives information about a range of scales at a fixed
	time of observation, rather than
	a fixed scale at a range of final times.
	The past-defined objects required
	for the latter
	are automatically provided by the forwards formalism, meaning
	that multiple integrations are not required.

	If the time of observation is known then the backwards
	formalism gives an efficient means to treat multiple
	scales at once.
	
	To extend the backwards formalism to the
	trispectrum, one needs to separate the
	observables $\tauNL$ and $\gNL$.
	For this purpose
	$N_{ij}$ and $N_{ijk}$ themselves would
	be required.
	Therefore,
	given the potential utility of this method, we
	conclude by extending it to include a 
	backwards evolution equation for $N_{IJ}$.
	As for the spectrum, this can be used
	to obtain information
	about $\fNL$ and $\tauNL$ over a range of scales
	at a fixed time of observation.
	It still requires
	the solution for only a two-index object. The transport equation for 
	$N_{IJ}$ can be shown to be
	\begin{equation}
		\frac{\d N_{JK}}{\d N} = -u_{IJK} N_I - u_{IJ} N_{IK} - u_{IK} N_{IJ} .
	\end{equation}
	This is to be solved backwards from the final hypersurface where
	$N_{IJ}$ is equal to $N_{\alpha \beta}$.
	
	\end{fmffile}

	\bibliography{paper}

\end{document}